\documentclass[letterpaper,aps,floatfix,twocolumn,tightenlines,amsmath,amssymb]{revtex4-1}

\usepackage{graphicx}
\usepackage{amssymb}
\usepackage{color}
\usepackage{psfrag}
\usepackage{ifsym}
\usepackage{epstopdf}
\usepackage{dsfont}
\usepackage{amsmath}
\usepackage{multirow} 
\usepackage{paralist}
\usepackage{xfrac}

\usepackage{amsthm}

\newcommand{\bra}[1] {\langle #1 |}
\newcommand{\ket}[1] {| #1 \rangle}

\newcommand{\ketbra}[1]{ | #1 \rangle\!\langle #1 |}

\newcommand{\ve} {\varepsilon}
\newcommand{\Tr} {\operatorname{Tr}}

\newcommand{\moy}[1]{\langle #1 \rangle}

\newcommand{\one}{\leavevmode\hbox{\small1\normalsize\kern-.33em1}}

\newcommand{\ba}{\begin{eqnarray}}
\newcommand{\ea}{\end{eqnarray}}

\begin{document}
\title{Joint quantum measurements with minimum uncertainty}

\author{Martin Ringbauer$^{1,2}$, Devon N. Biggerstaff$^{1,2}$, Matthew A. Broome$^{1,2}$, Alessandro Fedrizzi$^{1,2}$, Cyril Branciard$^{1}$ and Andrew G. White$^{1,2}$}
\affiliation{$^1$Centre for Engineered Quantum Systems, $^{2}$Centre for Quantum Computer and Communication Technology, School of Mathematics and Physics, University of Queensland, Brisbane,   QLD 4072, Australia}

\begin{abstract}
Quantum physics constrains the accuracy of joint measurements of incompatible observables. Here we test tight measurement-uncertainty relations using single photons. We implement two independent, idealized uncertainty-estimation methods, the 3-state method and the weak-measurement method, and adapt them to realistic experimental conditions. Exceptional quantum state fidelities of up to 0.99998(6) allow us to verge upon the fundamental limits of measurement uncertainty.   
\end{abstract}

\maketitle

Measurement---assigning a number to a property of a physical system---is the keystone of the natural sciences. Our belief in perfect measurement precision was shattered by the paradigm shift heralded by quantum physics almost a century ago. It is perhaps surprising that even today active debate persists over the fundamental limits on measurement imposed by quantum theory.

At the heart of this debate is Heisenberg's uncertainty principle~\cite{Heisenberg:1927ul}, which encompasses at least three distinct statements about the limitations on preparation and measurement of physical systems~\cite{Busch:2007fk}: 
\begin{inparaenum}[(\bgroup\itshape i\egroup )]
\item a system cannot be prepared such that a pair of non-commuting observables (\emph{e.g.}\ position--momentum) are arbitrarily well defined; 
\item such a pair of observables cannot be jointly measured with arbitrary accuracy; and
\item measuring one of these observables to a given accuracy disturbs the other accordingly.
\end{inparaenum} 

The preparation uncertainty (\emph{i}) was quantified rigorously by Kennard as~\cite{kennard1927qmb}
\begin{equation}
\Delta x \, \Delta p \ \geq \ \hbar/2\ ,
\label{eq:kennard}
\end{equation}
where $\Delta x$ and $\Delta p$ are the standard deviations of the position and momentum distributions of the prepared quantum system, respectively. 
For measurement uncertainty (\emph{ii}) and (\emph{iii}), the corresponding quantities of interest are the measurement inaccuracies $\varepsilon$ and disturbances $\eta$. In his original paper~\cite{Heisenberg:1927ul}, Heisenberg argued that the product of $\varepsilon_x$ and $\eta_p$ should obey a similar bound to~\eqref{eq:kennard} in a measurement-disturbance scenario; however, a formal proof was long lacking. Recently Busch \emph{et al.} provided such a proof for a relation of the form $\varepsilon_x\eta_p \geq \hbar/2$~\cite{busch2013phe}.

However, there has been controversy on whether such a relation holds in full generality~\cite{Ballentine:1970kx,ozawa2003pch,busch2013phe,rozema2013ndd,Ozawa:2013fk,dressel2013}. The point of contention is the choice of exact definitions for the measurement inaccuracies $\varepsilon$ and disturbances $\eta$. 
In their derivation, Busch \emph{et al.}~\cite{busch2013phe} independently maximized the inaccuracies and disturbances over all possible quantum states for a given measurement apparatus;
hence, their inaccuracies and disturbances are in general defined for \emph{different} states. When both quantities are defined on the \emph{same} quantum state, the relation $\varepsilon_x\eta_p \geq \hbar/2$ does not necessarily hold~\cite{Ballentine:1970kx,ozawa2003pch}. 

Following this observation, Ozawa~\cite{ozawa2003uvr,ozawa2004urj} and Hall~\cite{hall2004pih} derived new relations for the joint-measurement and the measurement-disturbance scenarios, for any pair of observables. These were recently tested experimentally with neutronic and photonic qubits~\cite{erhart2012edu,rozema2012vhm,Weston:2013fk,Sulyok:2013fk,Baek:2013ys}, demonstrating violation of a generalization of the above relation. Although universally valid, neither Ozawa's nor Hall's relations are optimal; Branciard improved these and derived \emph{tight} relations quantifying the \emph{optimal} trade-off between inaccuracies in approximate joint measurements and between inaccuracy and disturbance~\cite{branciard2013ete}, for the definitions of $\ve$ and $\eta$ used by Ozawa and Hall.

\begin{figure}[!h]
  \begin{center}
\includegraphics[width=\columnwidth]{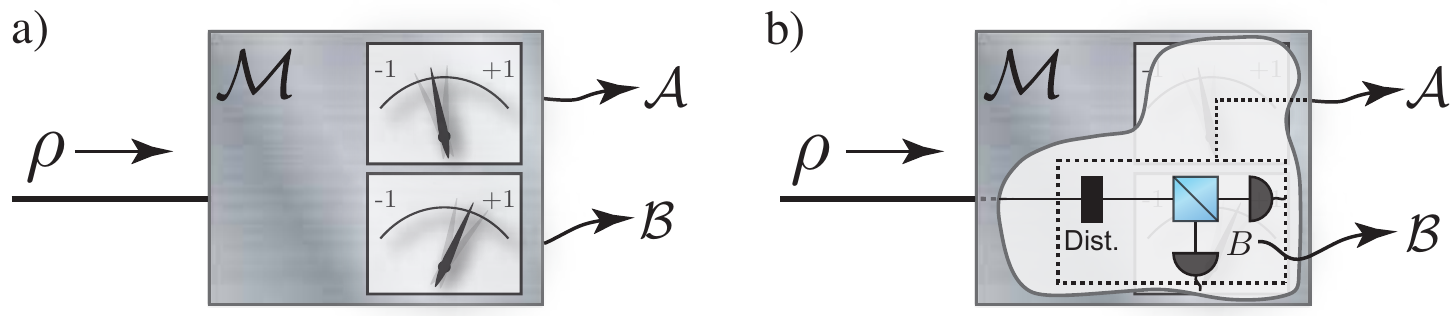}
  \end{center}
  \vspace{-4mm}
\caption{(a) The approximate joint measurement scenario: a quantum state $\rho$ is subjected to a measurement ${\mathcal M}$, from which observables ${\mathcal A}$ and ${\mathcal B}$ are extracted to approximate two incompatible observables $A$ and $B$, respectively.
(b)~In our implementation, an actual measurement of $B$ is performed after $\rho$ is disturbed (so as to also obtain an approximation of $A$). By opening the black box ${\mathcal M}$, our experiment can also be interpreted---when ${\mathcal B}$ is directly extracted from the disturbed measurement of $B$---as implementing the measurement-disturbance scenario. Note that this scenario requires $B$ and ${\mathcal B}$ to have the same spectrum.}
  \label{fig:Motivation}
\end{figure}

Here, we test Branciard's new relations by performing approximate joint measurements of incompatible polarization observables on single photons, see Fig.~\ref{fig:Motivation}(a). We verify that we can get close to saturating these relations in practice. Although framed within the joint-measurement scenario, our analysis also applies to the measurement-disturbance scenario, illustrated in Fig.~\ref{fig:Motivation}(b), in which case the inaccuracy $\ve_\mathcal{B}$ can be interpreted as the disturbance $\eta_\mathcal{B}$ on $B$.

We use two independent methods for estimating inaccuracies and disturbances experimentally: the 3-state method~\cite{ozawa2004urn} and the weak-measurement method~\cite{lund2010mmd}.
The 3-state method requires the preparation of multiple input states. The weak-measurement method, in turn, more closely resembles the classical approach for measuring inaccuracies, but comes at the cost of a more challenging experiment.
Crucially, both methods were defined under ideal conditions which are unattainable in practice. We therefore extend the respective estimation procedures to account for experimental imperfections---a step that has previously not received sufficient attention.

\paragraph{Theoretical framework.---}
Let $\rho$ denote a quantum state, and let $A$ and $B$ be two observables. The (in)compatibility of $A$ and $B$ when measured on $\rho$ is quantified by the parameter $C_{\!AB}{=}\frac{1}{2i} \Tr[(AB{-}BA){\cdot}\rho]$: whenever $C_{AB}{\neq}0$, they do not commute and cannot be jointly measured on $\rho$. However, one may still approximate their joint measurement using an observable ${\mathcal M}$ (or, more generally, a positive operator-valued measure (POVM) $\mathbb{M}$ \cite{Note1}) and defining approximations ${\mathcal A}{=}f({\mathcal M})$ and ${\mathcal B}{=}g({\mathcal M})$~\cite{Note3}, see Fig.~\ref{fig:Motivation}. Specifically, for an outcome $m$ of ${\mathcal M}$ and real-valued functions $f$ and $g$, one outputs $f(m)$ to approximate the measurement of $A$, and $g(m)$ to approximate the measurement of $B$.
Following Ozawa~\cite{ozawa2003uvr,ozawa2004urj}, one can quantify the inaccuracies of these approximations by the root-mean-square errors
\ba
\ve_{\mathcal A} = \Tr [({\mathcal A}{-}A)^2 \!\cdot\! \rho]^{1/2}, \quad
\ve_{\mathcal B} = \Tr [({\mathcal B}{-}B)^2 \!\cdot\! \rho]^{1/2}. \quad \label{eq:def_epsilons}
\ea

Branciard showed in~\cite{branciard2013ete} that, for any approximate joint measurement,
the above definitions of $\ve_{\mathcal A}$ and $\ve_{\mathcal B}$ satisfy the uncertainty relation for approximate joint measurements:
\begin{subequations}
  \begin{equation}
   \Delta B^2\ve_{\!\mathcal A}^2+\Delta A^2 \ve_{\!\mathcal B}^2+2 \sqrt{\Delta A^2 \Delta B^2{-}C_{\!AB}^2} \, \ve_{\!\mathcal A} \, \ve_{\!\mathcal B} \ {\geq} \ C_{\!AB}^2,
\label{eq:ETR}
\end{equation}
where $\Delta A{=}(\Tr [A^2 \rho]{-}\Tr [A \rho]^2 )^{1/2}$ and $\Delta B{=}(\Tr [B^2 \rho]-\Tr [B \rho]^2 )^{1/2}$ are the standard deviations of $A$ and $B$ on the state $\rho$. Furthermore, when $\rho$ is pure, this relation is \emph{tight}~\cite{branciard2013ete}: it quantifies the \emph{optimal} trade-off in the inaccuracies of the approximate measurements $\mathcal A$ and $\mathcal B$.

Interestingly, saturating Eq.~\eqref{eq:ETR} may require the approximate observables $\mathcal A$ and $\mathcal B$ to have different spectra from $A$ and $B$---\emph{i.e.}\ the optimal output values $f(m)$ and $g(m)$ may not be eigenvalues of $A$ and $B$.
One may nevertheless want to impose that the approximations ${\mathcal A}$ and/or ${\mathcal B}$ have the same spectrum as $A$ and $B$: this assumption is natural for ${\mathcal B}$ in a measurement-disturbance scenario, where ${\mathcal B}$ corresponds to an actual measurement of $B$ after $\rho$ is disturbed, see Fig.~\ref{fig:Motivation}(b). With this restrictive same-spectrum assumption, one can in general derive stronger relations than~\eqref{eq:ETR}~\cite{branciard2013ete}. For instance, in the case of $\pm 1$-valued observables (such that $A^2{=}B^2{=}\one$), when also imposing ${\mathcal A}^2{=}A^2{=}\one$ and/or ${\mathcal B}^2{=}B^2{=}\one$, relation~\eqref{eq:ETR} can be strengthened as follows~\cite{Branciard:ip}:
\begin{equation}
\textrm{Eq.}\eqref{eq:ETR}\; \begin{cases} 
\ve_{\mathcal{A}} \; &\hspace{-1em}\to\;  \min[1{-}(1{-}\ve_{\mathcal{A}}^2/2)^2, \, \Delta {A}^2]^{1/2}   \\
\ve_{\mathcal{B}} \; &\hspace{-1em}\to\;  \min[1{-}(1{-}\ve_{\mathcal{B}}^2/2)^2, \, \Delta {B}^2]^{1/2}
\end{cases}\ ,
\label{eq:EDR}
\end{equation}
\end{subequations}
where the replacement is made for the observable(s) on which the same-spectrum assumption is imposed. This relation generalizes the ``error-disturbance relation'' introduced in~\cite{branciard2013ete}. As well as testing \eqref{eq:ETR}, we also tested \eqref{eq:EDR} with the same-spectrum assumption imposed for both ${\mathcal A}$ and ${\mathcal B}$. The case where it is only imposed on ${\mathcal B}$ is presented in the Supplemental Material~\ref{Sec:Supp1}.

\begin{figure}[b]
  \begin{center}
\includegraphics[width=.9\columnwidth]{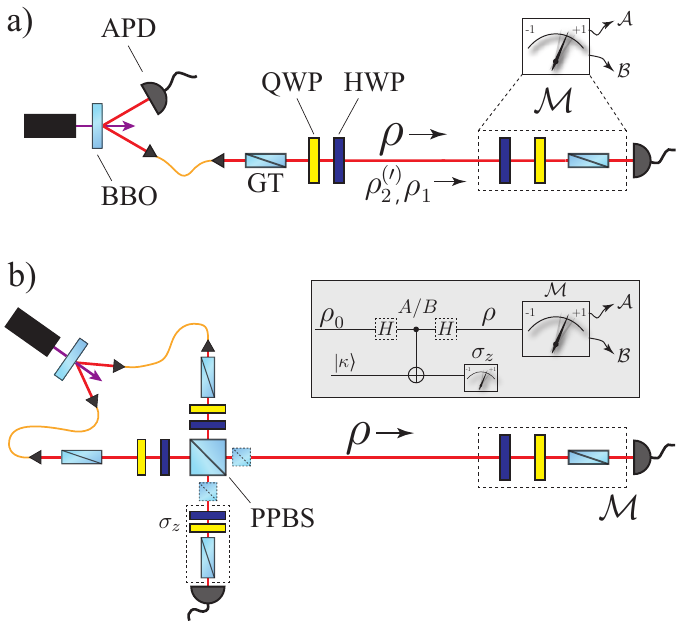}
  \end{center}
  \vspace{-3mm}
\caption{Experimental setup. \textbf{(a)} 3-state method. Single photons at a wavelength of $\lambda{=}820$~nm were produced in a non-collinear type-I spontaneous parametric down-conversion (SPDC) source using a $\beta$-barium-borate (BBO) non-linear crystal, pumped by a frequency-doubled fs-pulsed Ti:Sapph laser. State preparation was performed by Glan-Taylor polarizers (GT), a quarter-wave and a half-wave plate (QWP, HWP). A HWP at an angle of $\frac\theta 4$ and a polarizer implemented the measurement $\mathcal M$. The additional QWP between these elements was used for state tomography, with avalanche photodiodes (APD) used for detection. \textbf{(b)} Weak-measurement method. State preparation and final measurement were realized as in (a). The (semi-)weak measurement relied on a controlled-phase gate based on 2-photon interference at a partially polarizing beam splitter (PPBS) with nominal reflectivities of $R_H{=}0$ and $R_V{=}2/3$ for horizontal and vertical polarization, respectively~\cite{Langford2005}.
The required amplitude compensation (dashed PPBS) was performed via of pre-biased input states and all Hadamard gates were incorporated into either the state-preparation or the measurement wave plates. The corresponding circuit diagram is shown in the grey inset. Note that the \textsc{Cnot} gate is equivalent to a controlled-phase gate between Hadamard gates in the meter arm.}
  \label{fig:Setup}
\end{figure}

In order to test the relations~\eqref{eq:ETR} and~\eqref{eq:EDR} experimentally, one must determine the inaccuracies $\ve_{\mathcal A}$ and $\ve_{\mathcal B}$. If we expand Eq.~\eqref{eq:def_epsilons} (see \ref{Sec:Supp1}), $\ve_{\mathcal A}$ can be related to the measurement statistics of $\mathcal{A}$ on the states $\rho$, $A \rho A$ and $(\one{+}A) \rho (\one{+}A)/\|\cdot\|$, motivating the 3-state method~\cite{ozawa2004urn}. Alternatively, the weak-measurement method~\cite{lund2010mmd} links the definition of $\ve_\mathcal{A}$ to the joint probability distribution of an initial (semi-)weak measurement of $A$ followed by a measurement of $\mathcal{A}$. 
These two independent techniques allow us to estimate $\ve_{\mathcal A}$ and $\ve_{\mathcal B}$ without any assumptions about the actual measurement apparatus.

\paragraph{Experimental implementation.---}
Our experimental demonstration was performed with polarization-encoded qubits, see Fig.~\ref{fig:Setup}. Denoting the Pauli matrices $\sigma_{x,y,z}$, and their eigenstates $\ket{{\pm}x,y,z}$, we prepared $\rho{=}\ket{{+}y}\!\bra{{+}y}{=}\left(\one + \sigma_y\right)/2$ in the case of the 3-state method, and $\rho{=}\left(\one + \sqrt{1-\kappa^2} \sigma_y\right)/2$ for the weak-measurement method, where $\kappa \in [-1,1]$ quantifies the measurement strength. On these states we approximated the joint measurement of the incompatible observables $A{=}\sigma_x$ and $B{=}\sigma_z$.
For ideal states $\rho$, one finds $C_{\!AB}^2{=}1$ and $C_{\!AB}^2{=}1{-}\kappa^2$, respectively. 

The measurement apparatus implementing the joint approximation of $A$ and $B$ was chosen to perform a projective measurement ${\mathcal M}{=}\cos \theta \, \sigma_z{+}\sin \theta \, \sigma_x$ onto a direction in the $xz$-plane of the Bloch sphere.
In our experiment, this was realized by a half-wave plate 
and a polarizing prism which projected onto $\ket{{-}z}$, Fig.~\ref{fig:Setup}. The outcomes $m{=}{\pm}1$ of the measurement of $\mathcal{M}$ were then used to output some values $f(m)$ and $g(m)$.
These values were either chosen to minimize the inaccuracies $\ve_{\mathcal A}$ and $\ve_{\mathcal B}$, or set to $\pm 1$ in the case where the same-spectrum assumption was imposed, as discussed in~\ref{Sec:Supp1}.

For both experiments data was acquired for a series of settings $\theta \in [0,2\pi]$ of $\mathcal{M}$. We emphasize that in the calculations of $\ve_{\mathcal{A}}$ and $\ve_{\mathcal B}$ from either technique, we used neither the angle $\theta$ nor did we make any assumptions on the internal functioning of the measurement apparatus (e.g.\ that it implements a projective measurement): it is considered a black box that performs a fully general POVM with classical outputs $m{=}{\pm} 1$.

\begin{figure}[b]
  \begin{center}
 \includegraphics[width=.85\columnwidth]{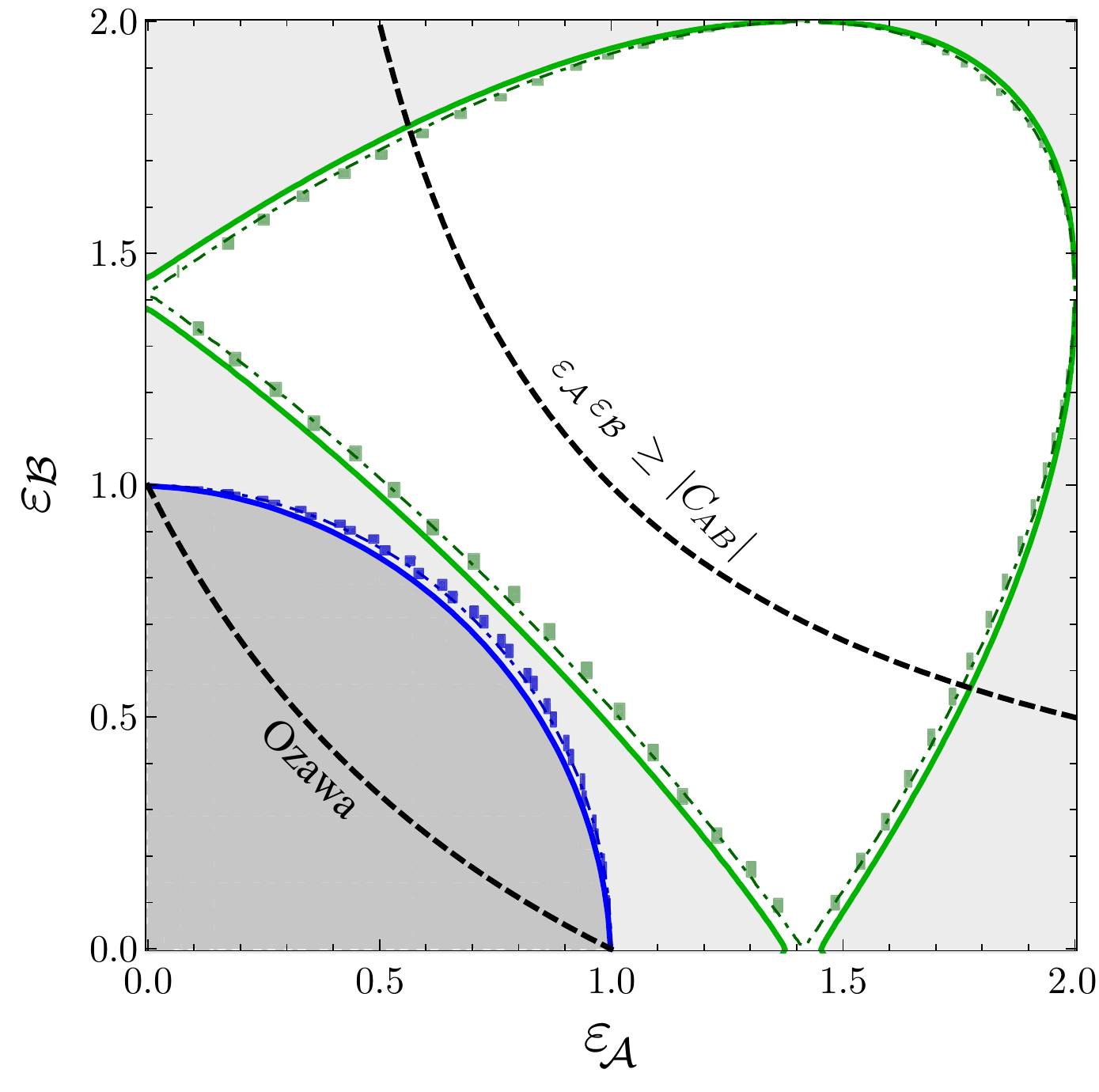}
  \end{center}
  \vspace{-5mm}
\caption{Experimental measurement inaccuracies, $\ve_{\mathcal A}$ vs. $\ve_{\mathcal B}$, characterized by the 3-state method. The blue rectangles represent the intervals of compatible values of $\ve_{\mathcal A}$ and $\ve_{\mathcal B}$ without the same-spectrum assumption. The solid blue curve corresponds to the bound imposed by the relation~\eqref{eq:ETR}, for the experimental values of $\Delta A, \Delta B$ and $C_{AB}$; the values below this bound are forbidden by quantum theory. The dot-dashed blue line is the bound imposed by~\eqref{eq:ETR} for the ideal case $\Delta A = \Delta B = C_{AB} = 1$.
The green rectangles and curves represent the corresponding data when the same-spectrum assumption is imposed on both ${\mathcal A}$ and ${\mathcal B}$, now invoking relation~\eqref{eq:EDR}; note that in contrast to~\eqref{eq:ETR}, this relation also \emph{upper}-bounds the values of $\ve_{\mathcal A}$ and $\ve_{\mathcal B}$.
For comparison, the black dashed curves indicate the bounds imposed by the relation $\ve_{\mathcal A} \, \ve_{\mathcal B} \geq |C_{\!AB}|$---which is violated by our data---and by Ozawa's relation~\cite{ozawa2003uvr,ozawa2004urj}---which is indeed satisfied, but cannot be saturated. Note, that the shown intervals include $1\sigma$ statistical errors obtained from Monte Carlo sampling assuming Poissonian photon-counting statistics.}
  \label{fig:tomo}
    \vspace{-2mm}
\end{figure}

\paragraph{The 3-state method.---}
For this method~\cite{ozawa2004urn}, in addition to the state $\rho{\simeq}\ket{{+}y}\!\bra{{+}y}$, we prepared the states $\rho_1{\simeq}A\rho A{\simeq}B\rho B{\simeq}\ket{{-}y}\!\bra{{-}y}$ and $\rho_2{\simeq}(\one+A) \rho (\one+A)/{\|\cdot\|}{\simeq}\ket{{+}x}\!\bra{{+}x}$ (respectively $\rho_2' \simeq (\one+B) \rho (\one +B)/\|\cdot\| \simeq \ket{{+}z}\!\bra{{+}z}$),  and characterized them using over-complete quantum state tomography~\cite{James2001}. The values of $\ve_{\mathcal A}$ (respectively $\ve_{\mathcal B}$) can then be estimated from the measurement statistics of ${\mathcal M}$ in these states~\cite{ozawa2004urn}, see~\ref{Sec:Supp1}.

The experimental setup is shown in Fig.~\ref{fig:Setup}a. 
The initial state $\rho$ was prepared on a heralded single photon with high quantum state fidelity $\mathcal{F}{=}0.999172(7)$ and purity $\mathcal{P}{=}0.99917(2)$, and gave $C_{\!AB}^2{=}0.99669(3)$. The states $\rho_1$ and $\rho_2$ were prepared with similar quality, see~\ref{Sec:Supp2}. 
Figure~\ref{fig:tomo} shows the results obtained for $\ve_\mathcal{A}$ and $\ve_\mathcal{B}$, with and without the same-spectrum assumption. We get very close to saturating relations~\eqref{eq:ETR} and~\eqref{eq:EDR}. 

Importantly, the equations used for obtaining $\ve_\mathcal{A}$ and $\ve_\mathcal{B}$ in the original 3-state proposal \cite{ozawa2004urn} assume perfect state preparation. Directly applying them to imperfect experimental states invalidates the derivation and leads to unreliable results.
In the Supplemental Material \ref{Sec:Supp1}, we extend the estimation procedure to realistic conditions. 
With careful characterization of the input states, our method yields finite intervals for $\ve_\mathcal{A}$ and $\ve_\mathcal{B}$ that are compatible with the experimental data, shown as shaded rectangles in Fig.~\ref{fig:tomo}.

\paragraph{The weak measurement method.---}
A weak measurement~\cite{Pryde:2005qf} aims at extracting partial information from a quantum system without disturbing it. It is typically realized by weakly coupling the system to a meter which is then subjected to a projective measurement.
In practice, weak measurements cannot be infinitely weak---they disturb the state 
onto which they are applied. In our case, the joint measurement of $A$ and $B$ is then approximated on the disturbed state $\rho$ \emph{after} the (semi-)weak measurement of $A$ and $B$, respectively. 
Note that this disturbance necessarily introduces mixture to $\rho$. As a consequence, it may no longer be possible to saturate~\eqref{eq:ETR} and~\eqref{eq:EDR}; in particular $C_{\!AB}^2$ will be decreased. As noted in~\cite{Weston:2013fk}, the weak-measurement method actually works for any measurement strength. However, to approach saturation, one should set it as low as possible.

The experimental weak-measurement setup is shown in Fig.~\ref{fig:Setup}(b). We realized the weak measurement using a non-deterministic linear-optical controlled-\textsc{not} (\textsc{Cnot}) gate \cite{Langford2005} acting on our initial signal qubit $\rho_0 = \ket{{+}y}\!\bra{{+}y} = (\one + \sigma_y) / 2$ and a meter qubit in the state  $\ket{\kappa}{=}\sqrt{\frac{1+\kappa}{2}}\ket{0}{+}\sqrt{\frac{1-\kappa}{2}}\ket{1}$, which determines the measurement strength $\kappa$~\cite{Pryde:2005qf}. 
The \textsc{cnot} gate alone, followed by a measurement of the meter qubit in the computational basis, enables the semi-weak measurement of $B{=}\sigma_z$, while the semi-weak measurement of $A{=}\sigma_x$ requires 2 additional Hadamard gates, see Fig.~\ref{fig:Setup}(b). In both cases, the initial state $\rho_0$ of the signal qubit is transformed to $\rho = (\one + \sqrt{1-\kappa^2} \sigma_y) / 2$. In practice, the disturbed states $\rho^A$ and $\rho^B$ after the weak measurements of $A$ and $B$, respectively, necessarily differ slightly. To account for that, we simply defined $\rho$---which enters the definitions of $\Delta A, \Delta B, \ve_\mathcal{A}, \ve_\mathcal{B}$ and $C_{\!AB}$ in relations~\eqref{eq:ETR} and~\eqref{eq:EDR}---to be the averaged disturbed state $\rho = \frac{1}{2}(\rho^A{+}\rho^B)$.

We characterized the quality of our gate operation using quantum process tomography~\cite{OBrien2004a}, obtaining a process fidelity of $\mathcal{F}_p{=}0.964(1)$.
We further measured a state fidelity of $\mathcal{F}{=}0.99998(6)$ of the average disturbed state $\rho$, with a reduced purity of $\mathcal{P}{=}0.964(1)$, corresponding to an average value of $\kappa=-0.262(4)$, for which we obtain $C_{\!AB}^2{=}0.928(2)$. For the 2-qubit states $\rho_{12}^A$ and $\rho_{12}^B$ after the interaction (corresponding to the semi-weak measurements of $A$ and $B$, respectively), we find fidelities of $\mathcal{F}{=0.9938(6)}$ and $\mathcal{F}{=}0.9958(3)$.
More details on the quality of the prepared states, the used measurement apparatus, and a full error analysis can be found in~\ref{Sec:Supp2}. 

\begin{figure}
  \begin{center}
 \includegraphics[width=.85\columnwidth]{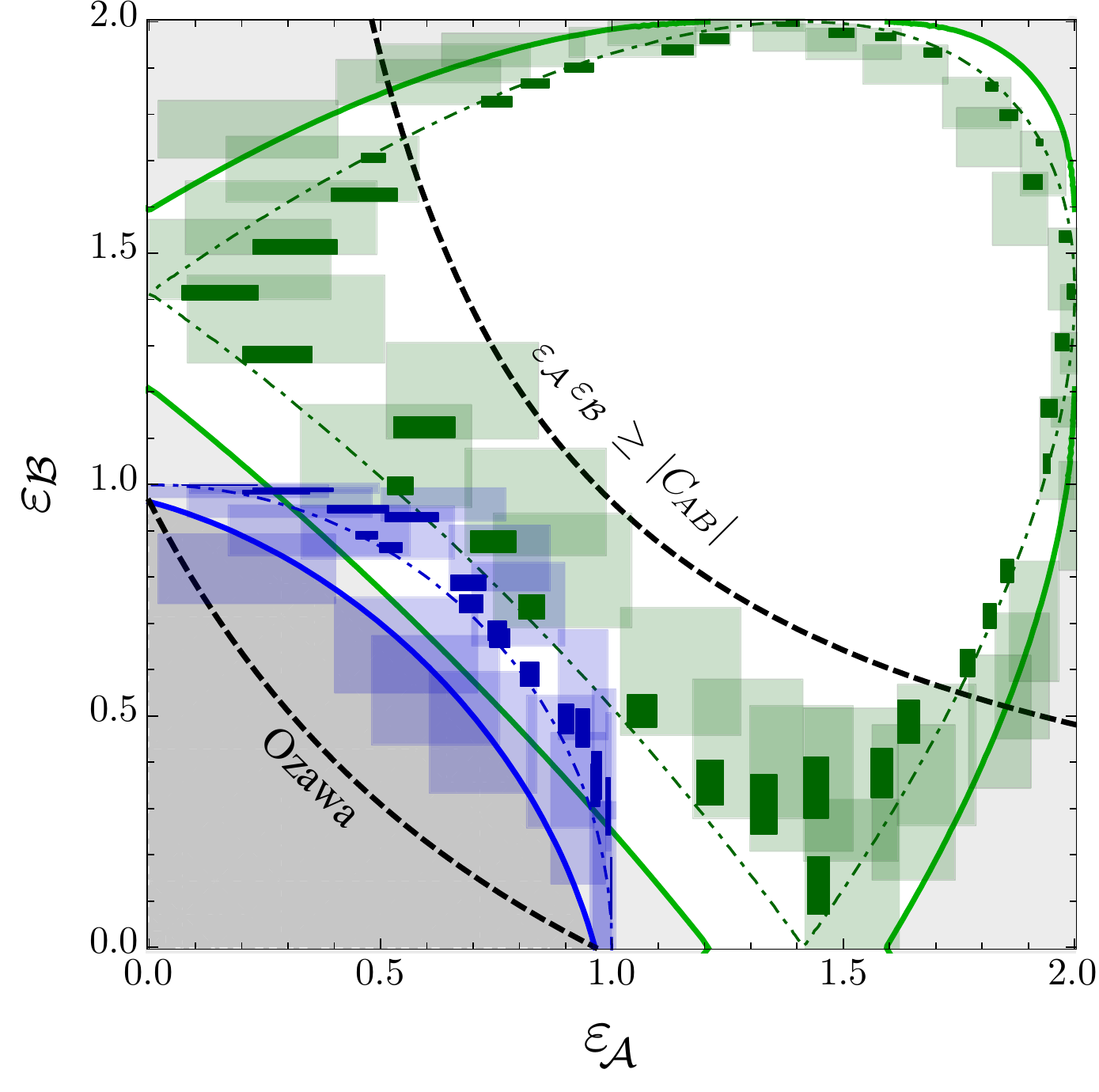}
   \vspace{-5mm}
  \end{center}
\caption{Results obtained via the weak-measurement method, presented as in Fig.~\ref{fig:tomo}. The darker rectangles represent smaller intervals of compatible values obtained by using experimental data from both the semi-weak measurements of $A$ and $B$. All intervals include $1\sigma$ statistical errors.}
  \label{fig:weak_meas}
    \vspace{-4mm}
\end{figure}

The derivations in the original proposal~\cite{lund2010mmd} for this method require the (semi-)weak measurements to be perfect. As for the 3-state method, we extend the estimation procedure for $\ve_\mathcal{A}$ and $\ve_\mathcal{B}$ to account for realistic experimental implementations and obtain intervals of compatible values, which are shown as rectangles which include $1\sigma$ statistical errors in Fig.~\ref{fig:weak_meas}. Furthermore we find that, if we, instead of treating them separately, take into account experimental data from both weak measurements of $A$ and $B$, the size of these intervals can be significantly reduced. The corresponding smaller intervals, shown as dark rectangles in Fig.~\ref{fig:weak_meas}, are now dominated by statistical errors, see~\ref{Sec:Supp1} for details.

\paragraph{Discussion.---} Our results agree with the theoretical predictions in all cases under consideration, indicating that one can indeed come close to saturating the measurement uncertainty relations~\eqref{eq:ETR} and~\eqref{eq:EDR} in the experiment. Unsurprisingly, the ranges of compatible values determined for the weak-measurement method are significantly larger than those for the 3-state method. This is due to the experimentally more demanding two-photon interaction. Although we took great care in the preparation of the initial states as well as in the alignment of the optical setup, residual errors from imperfect optical components, non-optimal spatio-temporal mode overlap and Poissonian counting statistics decrease the quality of the final data.

To put our data into context with previously proposed measurement-uncertainty relations, Figs.~\ref{fig:tomo} and \ref{fig:weak_meas} also show the relation $\ve_{\mathcal A} \, \ve_{\mathcal B} \geq |C_{\!AB}|$, by Arthurs and Kelly~\cite{arthurs1965osm} and Arthurs and Goodman~\cite{arthurs1988qca}, and Ozawa's relation~\cite{ozawa2003uvr,ozawa2004urj}. While the latter is universally valid and indeed satisfied by our data (but not saturated, as it is not tight), the former relation only holds under some restrictive assumptions~\cite{ozawa2003pch,arthurs1965osm,arthurs1988qca} for our definitions of $\ve_{\mathcal A}, \ve_{\mathcal B}$, and is clearly violated by our data.

Testing the ultimate measurement-uncertainty limits is crucial for our understanding of  quantum measurements. The new relations introduced in \cite{branciard2013ete} for both the joint measurement and the measurement-disturbance scenarios, and our comprehensive extension to experimental implementations, could play a role in refining a wide range of measurement-based quantum information protocols, such as quantum control or error correction. In particular, as detailed in~\ref{Sec:Supp1}, the optimal choice of approximating functions $f(m)$ and $g(m)$ in a joint-measurement experiment may differ from the theoretical optimum in the presence of experimental imperfections. Our technique of optimizing these quantities to find the optimal trade-off between $\ve_\mathcal{A}$ and $\ve_\mathcal{B}$ could be used as a calibration step in high-precision weak-, or joint-measurement experiments.

\begin{acknowledgments}
During the preparation of this manuscript we became aware of a related work by F.\ Kaneda et al.~\cite{Kaneda2013}. We thank G.\ G.\ Gillett and M.\ P.\ Almeida for assistance with the experiment. This work was supported in part by the Centres for Engineered Quantum Systems (CE110001013) and for Quantum Computation and Communication Technology (CE110001027). AF, CB, and AGW acknowledge support through Australian Research Council Discovery Early Career Award DE130100240, a University of Queensland (UQ) Postdoctoral Research Fellowship, and a UQ Vice-Chancellor's Senior Research Fellowship, respectively.
\end{acknowledgments}

\onecolumngrid
\clearpage
\renewcommand{\theequation}{S\arabic{equation}}
\renewcommand{\thefigure}{S\arabic{figure}}
\renewcommand{\thetable}{\Roman{table}}
\renewcommand{\thesection}{S\Roman{section}}
\setcounter{equation}{0}
\setcounter{figure}{0}
\begin{center}
{\bf \large Supplemental Material}
\end{center}
\twocolumngrid
\section{Data analysis: calculation of the inaccuracies $\ve_{\mathcal A}, \ve_{\mathcal B}$}
\label{Sec:Supp1}

Here we describe the procedure for estimating the inaccuracies $\ve_{\mathcal A}$ in our experiment, using both the 3-state and the weak measurement method, and in both 
the case where the same-spectrum is imposed, and the case where it is not. Importantly, we explain how to account for unavoidable experimental imperfections. The estimation of $\ve_{\mathcal B}$ is similar.

\medskip

Recall that $\rho$ denotes the 1-qubit state on which the joint measurement of the two incompatible observables $A$ and $B$ is approximated. In our case, $A$ and $B$ are $\pm 1$-valued, traceless qubit observables given by $A = \hat a \cdot \vec \sigma$ and $B = \hat b \cdot \vec \sigma$, where $\hat a$ and $\hat b$ are unit vectors on the Bloch sphere and $\vec \sigma{=}(\sigma_x, \sigma_y, \sigma_z)$.
The actual measurement apparatus is considered a black box which acts on qubits and outputs some binary values $m=\pm 1$. In full generality, we describe it as a 2-outcome POVM $\mathbb{M} = \{{\operatorname M}_m\} = \{{\operatorname M}_+, {\operatorname M}_-\}$ whose two elements ${\operatorname M}_\pm$ are 2$\times$2 non-negative hermitian operators which sum to the identity. 
They can be written as ${\operatorname M}_\pm = \frac{\one \pm (\mu \one + \vec m \cdot \vec \sigma)}{2}$, with $\mu \in \mathbb{R}$ and $\vec m = (m_x, m_y, m_z)$ a vector in the Bloch sphere, such that $|\mu| + ||\vec m|| \leq 1$. We also define ${\mathcal M} = {\operatorname M}_+ - {\operatorname M}_- = \mu \one + \vec m \cdot \vec \sigma$; when the POVM $\mathbb{M}$ is a projective measurement, ${\mathcal M}$ is the corresponding observable.

The outputs $m$ are used to define the estimates $f(m)$ and $g(m)$ for $A$ and $B$, respectively. Generalizing the definition of Eq.~(2) from the main text, the inaccuracy (root-mean-square error) $\ve_{\mathcal A}$ is given by~\cite{hall2004pih,ozawa2004urj,branciard2013ete}:
\begin{alignat}{3}
\varepsilon_{\mathcal A}^2 &= && {\sum}_m  \Tr  \left[\left(A-f(m)\one\right){\operatorname M}_m\left(A-f(m)\one\right) \rho \right] \nonumber \\
&=&\ 1 &+ {\sum}_m f(m)^2 \Tr  \left[{\operatorname M}_m \rho \right] \nonumber \\[-1mm]
&  &&- 2 {\sum}_m f(m) \operatorname{Re} \left[ \Tr  \left[{\operatorname M}_m A \rho \right] \right]  ,
\label{eq:epsA2}
\end{alignat}
where we used $A^2 = \one$ and $\Tr [A^2 \rho] = 1$ for a $\pm 1$-valued observable $A$.

Note that $\Tr\left[{\operatorname M}_m \rho \right]$ is simply the probability of outcome $m$, which can be directly estimated experimentally. What remains is to estimate the quantities $\operatorname{Re} \big[{\Tr}[{\operatorname M}_m A \rho] \big]$, or equivalently, in our case where $\mathbb{M}$ only has binary outputs,
\ba
\alpha_{\mathcal M} \ := \ \operatorname{Re} \big[ \Tr  [{\mathcal M} A \rho] \big] \ = \ \mu \, \moy{A}_\rho + \vec m \cdot \hat a \; , \label{eq:def_alphaM}
\ea
such that $\operatorname{Re} \big[ \Tr  [{\operatorname M}_\pm A \rho] \big] = \frac{\moy{A}_\rho \pm \alpha_{\mathcal M}}{2}$.
We employ two previously-suggested techniques to access these quantities experimentally: the 3-state method~\cite{ozawa2004urn} and the weak measurement method~\cite{lund2010mmd}.

\subsection{Estimating $\alpha_{\mathcal M}$ using the 3-state method}

The 3-state method allows in principle to estimate $\alpha_{\mathcal M}$ from a combination of the expectation values of $\mathbb{M}$ (or ${\cal M}$) in the 3 states $\rho$, $A \rho A$ and $(\one{+}A) \rho (\one{+}A)/\|(\one{+}A) \rho (\one{+}A)\|$. This becomes clear when rewriting $\operatorname{Re} \big[ \Tr  [{\mathcal M} A \rho] \big]$ in the form~\cite{ozawa2004urn}
\ba
\operatorname{Re} \big[ \Tr  [{\mathcal M} A \rho] \big] &=& \frac{1}{2} \Big[ \Tr  \big[{\mathcal M} (\one{+}A) \rho (\one{+}A) \big] \nonumber \\[-2mm]
&& - \Tr  \big[{\mathcal M} A \rho A \big] - \Tr  \big[{\mathcal M} \rho \big] \Big]. \quad
\label{eq:alphaM_tomo}
\ea

Due to experimental imperfections, however, it is not possible in practice to perfectly prepare the states $A \rho A$ and $(\one{+}A) \rho (\one{+}A)/\|(\one{+}A) \rho (\one{+}A)\|$, and calculate $\alpha_{\mathcal M}$ precisely as in Eq.~\eqref{eq:alphaM_tomo}. Nevertheless, performing measurements on two (well-characterized) additional states $\rho_1$ ($\simeq A \rho A$) and $\rho_2$ ($\simeq (\one{+}A) \rho (\one{+}A)$, up to normalization) allows one to restrict the possible values of $\alpha_{\mathcal M}$ to a small interval.

One can experimentally estimate the average values of $\mathbb{M}$ in the 3 states $\rho$, $\rho_1$ and $\rho_2$, which we denote $\moy{{\mathcal M}}_{\rho}^{exp}$, $\moy{{\mathcal M}}_{\rho_1}^{exp}$ and $\moy{{\mathcal M}}_{\rho_2}^{exp}$. Representing the three states by vectors $\vec \rho$, $\vec \rho_1$ and $\vec \rho_2$ in the Bloch sphere (such that $\rho = \frac{\one + \vec \rho \cdot \vec \sigma}{2}$, etc.), these expectation values are given by~\cite{NoteS1} $\moy{{\mathcal M}}_{\rho} = \mu + \vec m \cdot \vec \rho$, etc. To be compatible with the experimental observations, $(\mu, \vec m)$ must therefore be in the set
\ba
S_{\rho,\rho_1,\rho_2}^{exp} &=& \left\{ (\mu, \vec m) \ \left|
\begin{array}{l}
|\mu| + ||\vec m|| \leq 1 \ \text{ and } \\[2mm]
\left\{
\begin{array}{lcl}
\mu + \vec m \cdot \vec \rho &=& \moy{{\mathcal M}}_{\rho}^{exp}  \\[1mm]
\mu + \vec m \cdot \vec \rho_1 &=& \moy{{\mathcal M}}_{\rho_1}^{exp}  \\[1mm]
\mu + \vec m \cdot \vec \rho_2 &=& \moy{{\mathcal M}}_{\rho_2}^{exp}
\end{array}
\right.
\end{array}
\right.
\right\} . \qquad \label{eq:def_Srhos}
\ea
Using~\eqref{eq:def_alphaM}, we thus conclude that $\alpha_{\mathcal M}{\in}[\alpha_{\mathcal M}^{min}, \alpha_{\mathcal M}^{max}]$ with
\ba
\alpha_{\mathcal M}^{min (max)} = \underset{(\mu, \vec m) \in S_{\rho,\rho_1,\rho_2}^{exp}}{\min (\max)} \big[ \, \mu \, \moy{A}_\rho + \vec m \cdot \hat a \, \big] . \quad
\ea
Given the small dimensions of this optimization problem in our case, it can trustfully be solved with standard algorithms~\cite{NoteS2}.
Note that the states $\rho, \rho_1$ and $\rho_2$ must be carefully characterized so that the constraints on $(\mu, \vec m)$ describing the set $S_{\rho,\rho_1,\rho_2}^{exp}$ in~\eqref{eq:def_Srhos} are sharply defined.

Therefore in practice the values of $\alpha_{\mathcal M}$ cannot be determined precisely via the 3-state method, but it is still possible to restrict them to some interval. The choice $\rho_1 \simeq A \rho A$ and $\rho_2 \simeq (\one{+}A) \rho (\one{+}A)/\|(\one{+}A) \rho (\one{+}A)\|$, motivated by Eq.~\eqref{eq:alphaM_tomo}, should ensure the range of possible values of $\alpha_{\mathcal M}$ to be small (i.e.\ $\alpha_{\mathcal M}^{min} \simeq \alpha_{\mathcal M}^{max}$). However, any choice for $\rho_1$ and $\rho_2$ would actually yield a finite interval for $\alpha_{\mathcal M}$. Moreover, to further restrict this interval one can also use more input states; a fourth input state (e.g.\ $\rho_2'$, used for the estimation of $\ve_{\mathcal B}$) can indeed allow one to precisely determine $\mu$ and $\vec m$, and therefore $\alpha_{\mathcal M}$ through~\eqref{eq:def_alphaM}. This technique then amounts to performing a tomography of the measurement apparatus.

\begin{figure}
  \begin{center}
\includegraphics[width=\columnwidth]{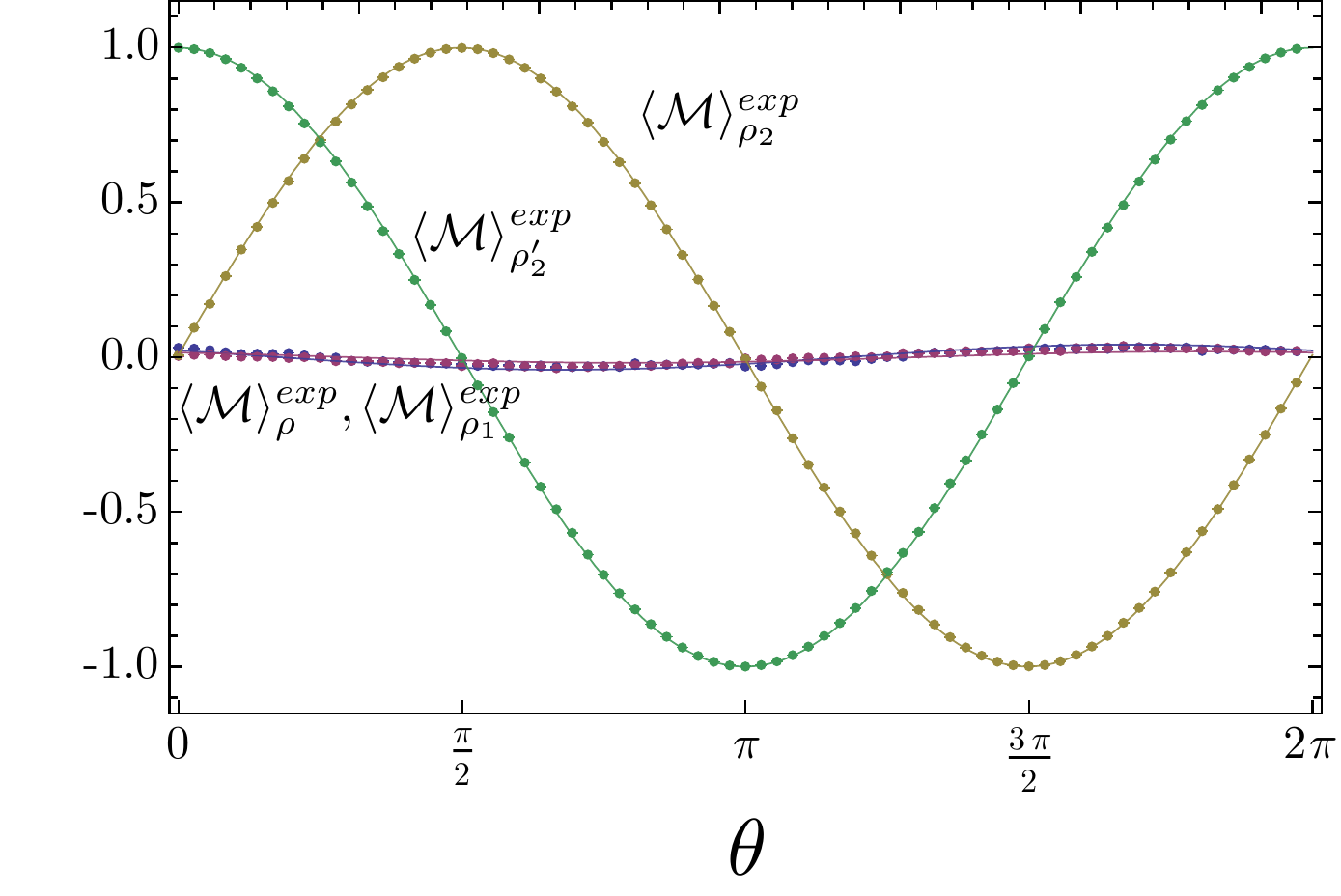}
  \end{center}
\caption{Measured expectation values used for the 3-state method for various orientations $\theta$ of the measurement apparatus. The solid curves represent theoretical predictions from the characterization of the prepared states. 
Errors from Poissonian counting statistics are too small to be visible.}
  \label{fig:Supp_ExpVal_Tomo}
\end{figure}

In our experiment, where $A=\sigma_x$ and $\rho \simeq \ket{{+}y}\!\bra{{+}y}$, we prepared the states $\rho$, $\rho_1 \simeq A \rho A \simeq \ket{{-}y}\!\bra{{-}y}$ and $\rho_2 \simeq (\one{+}A) \rho (\one{+}A)/\| \cdot \| \simeq \ket{{+}x}\!\bra{{+}x}$. The preparation and characterization of these states is detailed in section~\ref{sec:Supp_Exp_Tomo} below. For various orientations $\theta{\in}[0,2\pi ]$ of the measurement apparatus we estimated the expectation values $\moy{{\mathcal M}}_{\rho}^{exp}$, $\moy{{\mathcal M}}_{\rho_1}^{exp}$ and $\moy{{\mathcal M}}_{\rho_2}^{exp}$, shown in Fig.~\ref{fig:Supp_ExpVal_Tomo}. From the calculation detailed above we could then determine the intervals $[\alpha_{\mathcal M}^{min}, \alpha_{\mathcal M}^{max}]$.

\begin{figure}
  \begin{center}
\includegraphics[width=\columnwidth]{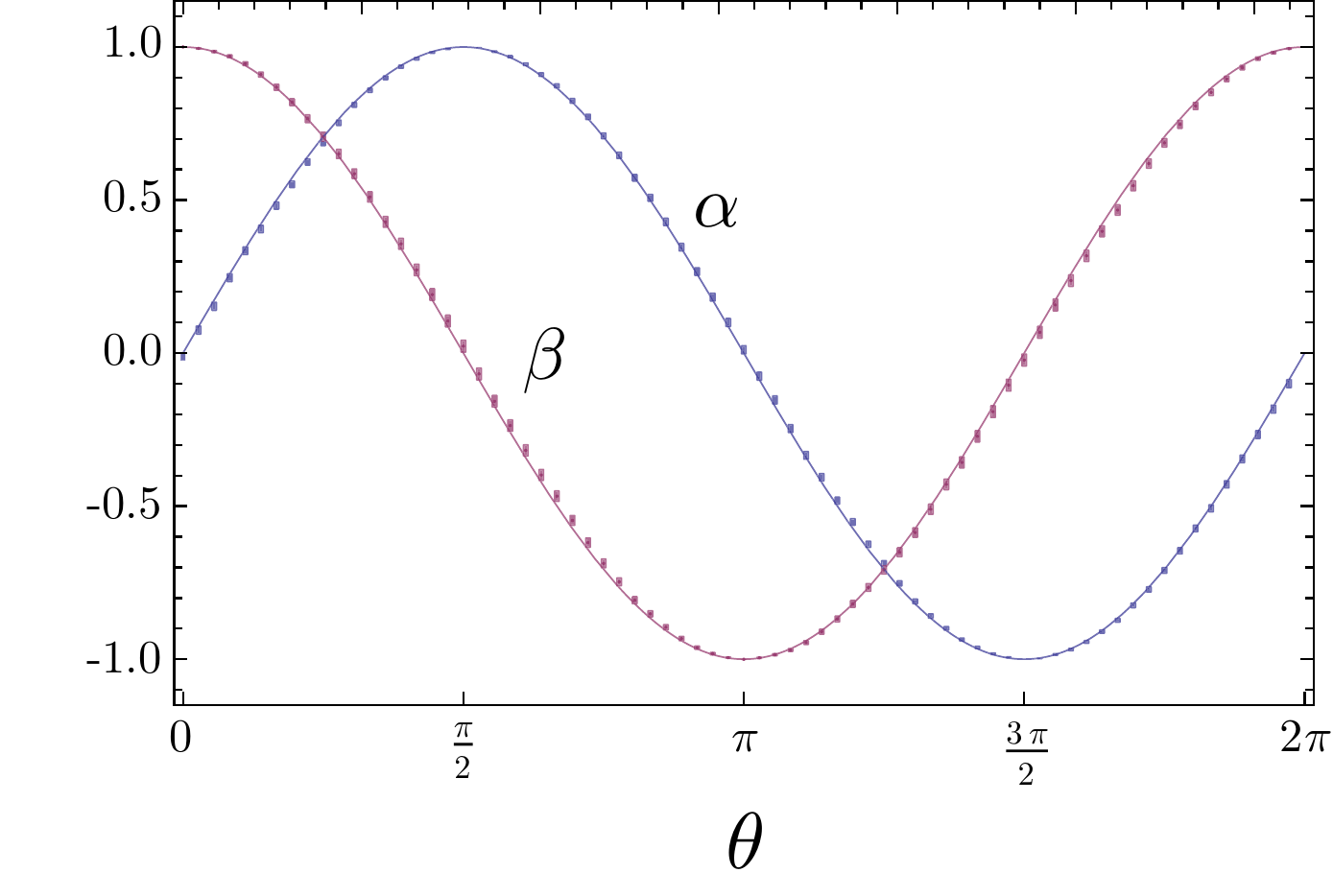}
  \end{center}
\caption{The intervals $[\alpha_{\mathcal M}^{min}, \alpha_{\mathcal M}^{max}]$ and $[\beta_{\mathcal M}^{min}, \beta_{\mathcal M}^{max}]$ obtained from the 3-state method for various orientations $\theta$ of the measurement apparatus are shown as shaded vertical line segments. We observe good agreement with the theoretical values determined from the characterization of the prepared states, shown as solid curves.}
  \label{fig:Supp_AlphaBeta_Tomo}
\end{figure}

In order to estimate the ranges $[\beta_{\cal M}^{min}, \beta_{\cal M}^{max}]$ of possible values for $\beta_{\cal M} = \text{Re} \big[ \text{Tr} [{\cal M} B \rho] \big]$ (with $B=\sigma_z$), we similarly prepared the states $\rho$, $\rho_1 \simeq B \rho B \simeq \ket{{-}y}\!\bra{{-}y} (\simeq A \rho A)$ and $\rho_2' \simeq (\one + B) \rho (\one + B)/\| \cdot \| \simeq \ket{{+}z}\!\bra{{+}z}$. We estimated the average values $\moy{{\cal M}}_{\rho}^{exp}$, $\moy{{\cal M}}_{\rho_1}^{exp}$ (the same as above) and $\moy{{\cal M}}_{\rho_2'}^{exp}$, see Fig.~\ref{fig:Supp_ExpVal_Tomo}. The intervals $[\alpha_{\mathcal M}^{min}, \alpha_{\mathcal M}^{max}]$ and $[\beta_{\cal M}^{min}, \beta_{\cal M}^{max}]$, including $1\sigma$ statistical errors obtained from Monte Carlo sampling of the Poissonian photon-counting statistics, are shown in Figure~\ref{fig:Supp_AlphaBeta_Tomo}.

\subsection{Estimating $\alpha_{\mathcal{M}}$ using the weak measurement method}
\label{Sec:Supp_Alpha_Weak}

The idea of the method proposed in~\cite{lund2010mmd} is to understand $\alpha_{\mathcal M} = \operatorname{Re} \big[ \Tr  [{\mathcal M} A \rho] \big]$ as a so-called weak-valued average. Specifically it is the average of the product $m\, a$ over the weak-valued joint probability distribution $P_{\rho}^{wv}\big(m,a\big)$, which represents the probability that an initial weak measurement of $A$ yields result $a$ and the final measurement $\mathbb{M}$ on $\rho$ yields outcome $m$~\cite{Wiseman2003}. This weak-valued joint probability distribution is given by
\ba
P_{\rho}^{wv}\big(m,a\big) &=& \operatorname{Re}\big[\Tr [{\operatorname M}_m \Pi_a^A \rho]\big] , 
\label{eq:Pwv}
\ea
where $\Pi_a^A$ is the projector corresponding to the eigenvalue $a$ of $A$ (such that $A = \Pi_{+1}^A - \Pi_{-1}^A$ in our case of a dichotomic observable, with $a = \pm 1$), and one obtains $\sum_{m,a} m \, a \, P_{\rho}^{wv}\big(m,a\big) = \alpha_{\mathcal M}$. Note that weak-valued joint probabilities can be negative, but are always normalized: $\sum_{m, a} P_{\rho}^{wv}\big(m,a\big) = 1$~\cite{Wiseman2003}.

The weak measurement method thus consists of performing a weak measurement of $A$ before the final measurement of $\mathcal{M}$, and estimating $\alpha_{\mathcal M}$ from the average of the product of outcomes $m \, a$. Crucially, as infinitely weak measurements cannot be implemented experimentally, the semi-weak measurement will necessarily perturb the state. As we shall see, this is actually not a problem if one takes the state $\rho$ (which enters the measurement apparatus and on which the inaccuracies $\ve_{\mathcal A}$ and $\ve_{\mathcal B}$ are defined) to be the perturbed state, after the weak measurement. Importantly, the semi-weak measurement need not be particularly weak: the general method works for any measurement strength (as also noted in~\cite{Weston:2013fk}). However, increasing the strength results in a more mixed state $\rho$, which in turn makes it harder to approach and saturate the bounds imposed by the measurement uncertainty relations~(3a) and~(3b) of the main text.

\medskip

In our case, a weak measurement of the observable $A = \sigma_x$ on the initial state $\rho_0 = \ket{{+}y}\!\bra{{+}y}$ is achieved through a sequence of Hadamard ($U_H$) and \textsc{cnot} ($U_{\textsc{cnot}}$) gates~\cite{NielsenChuang} acting on the joint 2-qubit state $\ket{{+}y}\otimes\ket{\kappa}$, with an ancillary state  $\ket{\kappa}{=}\sqrt{\frac{1+\kappa}{2}}\ket{0}{+}\sqrt{\frac{1-\kappa}{2}}\ket{1}$. After the unitary transformation $U_A = (U_H \otimes \one) \cdot U_{\textsc{cnot}} \cdot (U_H \otimes \one)$, the state is given by
\ba
\rho_{12}^{A,th} &=& U_A \cdot \big( \, \ket{{+}y}\!\bra{{+}y} \otimes \ket{\kappa}\!\bra{\kappa} \, \big) \cdot U_A^\dagger \nonumber \\
&=& \frac{1}{4} \Big( \one + \sqrt{1-\kappa^2} \, \sigma_y \otimes \one  + \sqrt{1-\kappa^2} \, \one \otimes \sigma_x \nonumber \\[-2mm]
&& \qquad + \sigma_y \otimes \sigma_x + \kappa \, \sigma_x \otimes \sigma_z -
 \kappa \, \sigma_z \otimes \sigma_y \Big) . \quad
\label{eq:rho12Ath}
\ea
Here subscript 1 denotes the system qubit and 2 the ancillary qubit, while the superscript $th$ indicates that these are ideal, theoretical states.
After the interaction, the first qubit is in the state
\ba
\rho_{1}^{A,th} &=& \Tr _2 \, \rho_{12}^{A,th} \ = \ \frac{1}{2} \Big( \one + \sqrt{1-\kappa^2} \, \sigma_y \Big) .
\label{eq:rho1Ath}
\ea

For this ancillary state $\ket{\kappa}$ and unitary $U_A$, a projective measurement of $\sigma_z$ on the ancillary qubit after the interaction, with outcomes denoted by $a = \pm 1$, effectively amounts to performing a POVM with elements $\Tr _2 [(U_A^\dagger \cdot (\one \otimes \frac{\one \pm \sigma_z}{2}) \cdot U_A) \cdot (\one \otimes \ket{\kappa}\!\bra{\kappa})] = \frac{\one \pm \kappa \sigma_x}{2}$ on the first qubit. This therefore implements a weak measurement of $A = \sigma_x$ on the first qubit, where $\kappa\in[-1,1]$ quantifies the strength of the measurement: the smaller $|\kappa|$, the weaker the measurement.
Recalling that ${\mathcal M} = \mu \one + \vec m \cdot \vec \sigma$, the average value of the product of outcomes $m \, a$ for the measurements on $\rho_{12}^{A,th}$ of Eq.~\eqref{eq:rho12Ath} is
\ba
\moy{{\mathcal M} \otimes \sigma_z}_{\rho_{12}^{A,th}} &=& \kappa \, m_x .
\label{kmx}
\ea

Similarly, in order to weakly measure $B = \sigma_z$ on $\rho_0$, one can prepare the same joint 2-qubit state $\ket{{+}y} \otimes \ket{\kappa}$ and apply a \textsc{Cnot} gate only (implementing $U_B = U_{\textsc{cnot}}$), after which the state is
\ba
\rho_{12}^{B,th} &=& U_B \cdot \big( \, \ket{{+}y}\!\bra{{+}y} \otimes \ket{\kappa}\!\bra{\kappa} \, \big) \cdot U_B^\dagger \nonumber \\
&=& \frac{1}{4} \Big( \one + \sqrt{1-\kappa^2} \, \sigma_y \otimes \one  + \sqrt{1-\kappa^2} \, \one \otimes \sigma_x \nonumber \\[-2mm]
&& \qquad + \sigma_y \otimes \sigma_x + \kappa \, \sigma_x \otimes \sigma_y +
 \kappa \, \sigma_z \otimes \sigma_z \Big) . \qquad
\label{eq:rho12Bth}
\ea
The state of the first qubit is accordingly
\ba
\rho_{1}^{B,th} &\!=\!& \Tr _2 \, \rho_{12}^{B,th} = \frac{1}{2} \Big( \one + \sqrt{1-\kappa^2} \, \sigma_y \Big) = \rho_{1}^{A,th}. \ \qquad
\label{eq:rho1Bth}
\ea
Hence, the average disturbed state on which the joint measurement of $A$ and $B$ is approximated is $\rho^{th} = \rho_{1}^{A,th} = \rho_{1}^{B,th}$.

A final projective measurement of $\sigma_z$ on the second qubit of $\rho_{12}^{B,th}$ (with outcome $b{=}{\pm}1$), now effectively implements a POVM $\{\frac{\one \pm \kappa \sigma_z}{2}\}$---a weak measurement of $B{=}\sigma_z$---on the first qubit. The expectation value of the product of outcomes $m \, b$, for the measurements on $\rho_{12}^{B,th}$ of Eq.~\eqref{eq:rho12Bth}, is
\ba
\moy{{\mathcal M} \otimes \sigma_z}_{\rho_{12}^{B,th}} &=& \kappa \, m_z .
\label{eq:kmz}
\ea

For $A{=}\sigma_x$, $B{=}\sigma_z$, using~\eqref{eq:def_alphaM} (with $\moy{A}_{\rho^{th}}{=}\moy{B}_{\rho^{th}}{=}0$), together with \eqref{kmx} and~\eqref{eq:kmz}, we find
\ba
\alpha_{\mathcal M} &=& m_x \ = \ \frac{ \moy{{\mathcal M} \otimes \sigma_z}_{\rho_{12}^{A,th}} }{\kappa}, \label{eq:alphaM_weak_meas} \\
\beta_{\mathcal M} &=& m_z \ = \ \frac{ \moy{{\mathcal M} \otimes \sigma_z}_{\rho_{12}^{B,th}} }{\kappa}.
\ea
Hence, the values $\alpha_{\mathcal M}$ and $\beta_{\mathcal M}$ can in principle be estimated directly from the experimentally accessible quantities $\moy{{\mathcal M} \otimes \sigma_z}_{\rho_{12}^{A,th}}$ and $\moy{{\mathcal M} \otimes \sigma_z}_{\rho_{12}^{B,th}}$ (this technique was first presented and implemented in Ref.~\cite{Pryde:2005qf}).

In every experimental implementation, however, the state preparation and the 2-qubit interactions will necessarily be imperfect, resulting in approximate states $\rho_{12}^{A} \simeq \rho_{12}^{A,th}$ and $\rho_{12}^{B} \simeq \rho_{12}^{B,th}$. Nevertheless, one can still use these imperfect states---provided they are carefully characterized---to restrict the possible values of $\alpha_{\mathcal M}$ and $\beta_{\mathcal M}$ to some small intervals, similarly as in the 3-state method.
Furthermore, $\rho_1^A$ and $\rho_1^B$ (with $\rho_1^{A/B}{=}\Tr _2 \, \rho_{12}^{A/B}$) will not be equal in general. Therefore, the state $\rho$---on which the joint measurement of $A$ and $B$ is approximated, and which enters the definition of the inaccuracies $\ve_{\mathcal A}$ and $\ve_{\mathcal B}$, of $\alpha_{\mathcal M}$ and $\beta_{\mathcal M}$, and of $\moy{A}_\rho$ and $\moy{B}_\rho$ in particular---will be taken as the average state $\rho = \frac{1}{2} (\rho_1^A + \rho_1^B)$.

Inspired by the theoretical analysis above, the value of $\alpha_{\mathcal M}$ can now be bounded using the expectation value {$\moy{{\mathcal M} \otimes \sigma_z}_{\rho_{12}^{A}}^{exp}$}. Note that one has experimentally also access to $\moy{{\mathcal M} \otimes \one}_{\rho_{12}^{A}}^{exp} = \moy{{\mathcal M}}_{\rho_{1}^{A}}^{exp}$; taking it into account will be useful to further restrict the possible values of $\alpha_{\mathcal M}$. Recalling that ${\mathcal M} = \mu \one + \vec m \cdot \vec \sigma$, the above expectation values take the form
\ba
\moy{{\mathcal M} \otimes \sigma_z}_{\rho_{12}^{A}} \ &=& \ \mu \, \rho_{12}^{A,1z} + \vec m \cdot \vec \rho_{12}^{A,\cdot z} \label{eq:ExpMz_equation}\\[1mm]
\quad \text{with } \rho_{12}^{A,1z} &=& \moy{\one \otimes \sigma_z}_{\rho_{12}^A} \nonumber \\
\quad \text{and } \vec \rho_{12}^{A,\cdot z} &=& (\moy{\sigma_x \!\otimes\! \sigma_z}_{\rho_{12}^A}, \moy{\sigma_y \!\otimes\! \sigma_z}_{\rho_{12}^A}, \moy{\sigma_z \!\otimes\! \sigma_z}_{\rho_{12}^A}), \nonumber
\ea
and $\moy{{\mathcal M} \otimes \one}_{\rho_{12}^{A}} = \mu + \vec m \cdot \vec \rho_1^A$, where $\vec \rho_1^A$ is the Bloch vector representing $\rho_1^A$.
In order to be compatible with the experimental observations, $(\mu, \vec m)$ must therefore be in the set
\ba
T_{\rho_{12}^{A}}^{exp} &=& \left\{ \! (\mu, \vec m) \left|
\begin{array}{l}
|\mu| + ||\vec m|| \leq 1 \ \text{ and } \\[2mm]
\left\{
\begin{array}{lcl}
\! \mu \, \rho_{12}^{A,1z} + \vec m \! \cdot \! \vec \rho_{12}^{A,\cdot z} &\!=\!& \moy{{\mathcal M} \! \otimes \! \sigma_z}_{\rho_{12}^{A}}^{exp}  \\[1mm]
\! \mu + \vec m \cdot \vec \rho_1^A &\!=\!& \moy{{\mathcal M}}_{\rho_{1}^{A}}^{exp}
\end{array}
\!\!\!\! \right.
\end{array}
\right.
\right\} \! . \nonumber \\
\label{eq:def_TrhoA}
\ea
Using~\eqref{eq:def_alphaM}, we can conclude that $\alpha_{\mathcal M}{\in}[\alpha_{\mathcal M}^{min}, \alpha_{\mathcal M}^{max}]$ with
\ba
\alpha_{\mathcal M}^{min (max)} = \underset{(\mu, \vec m) \in T_{\rho_{12}^{A}}^{exp}}{\min (\max)} \big[ \, \mu \, \moy{A}_\rho + \vec m \cdot \hat a \, \big] . \quad 
\label{eq:alpha_min_max_T}
\ea
Note again that a careful characterization of the state $\rho_{12}^A$ is essential for the constraints on $(\mu, \vec m)$, describing the set $T_{\rho_{12}^{A}}^{exp}$ in~\eqref{eq:def_TrhoA}, to be sharply defined. We indeed find that Eq.~\eqref{eq:ExpMz_equation}, in particular, is increasingly sensitive to experimental imperfections with decreasing measurement strength $|\kappa|$.

\begin{figure}
  \begin{center}
\includegraphics[width=\columnwidth]{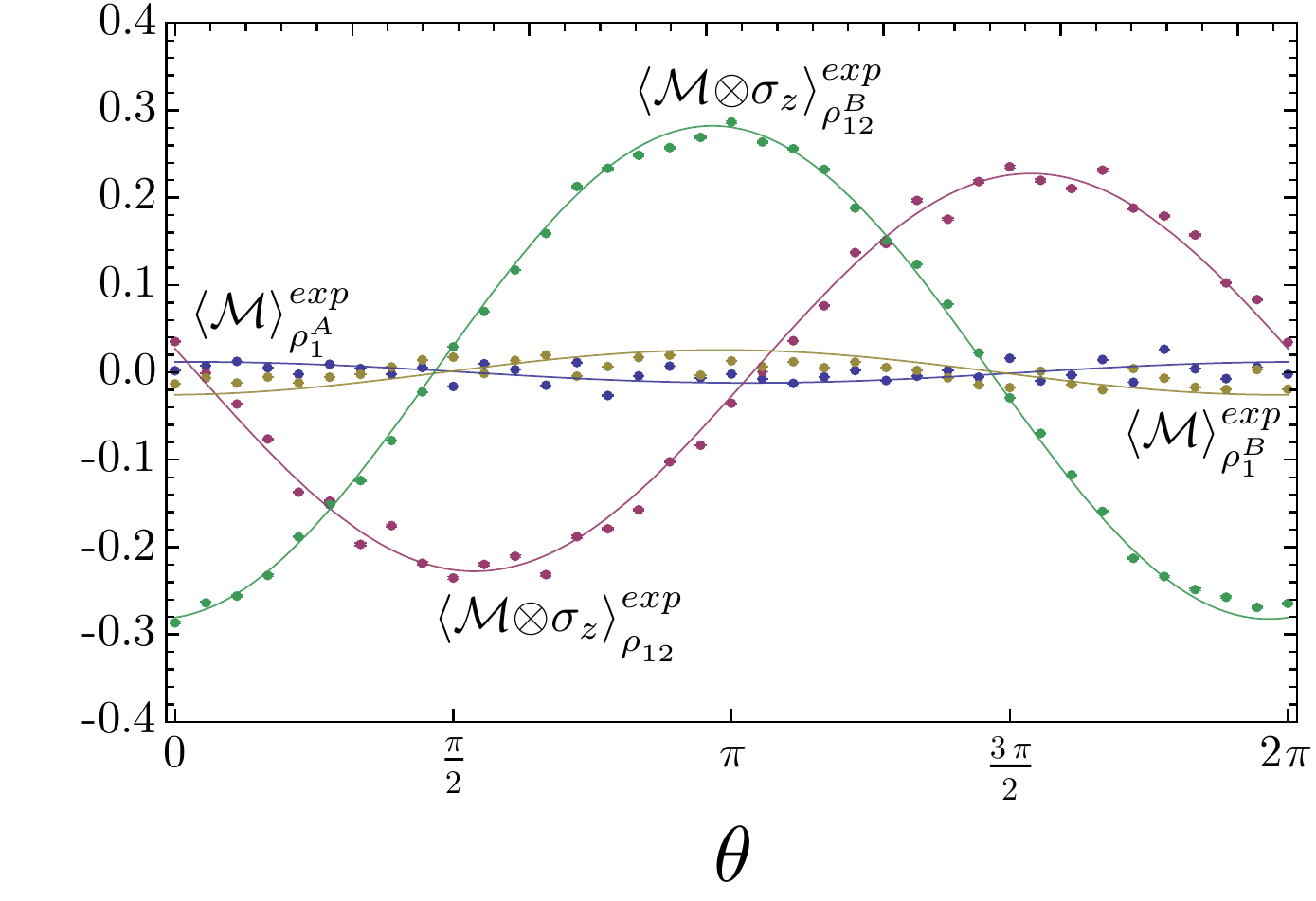}
  \end{center}
\caption{Measured expectation values used for the weak measurement method for various orientations $\theta$ of the measurement apparatus. The solid curves represent theoretical predictions from the characterization of the prepared states. The errors from the Poissonian counting statistics are too small to be visible.}
  \label{fig:Supp_ExpVal_Weak}
\end{figure}

\begin{figure}
  \begin{center}
\includegraphics[width=\columnwidth]{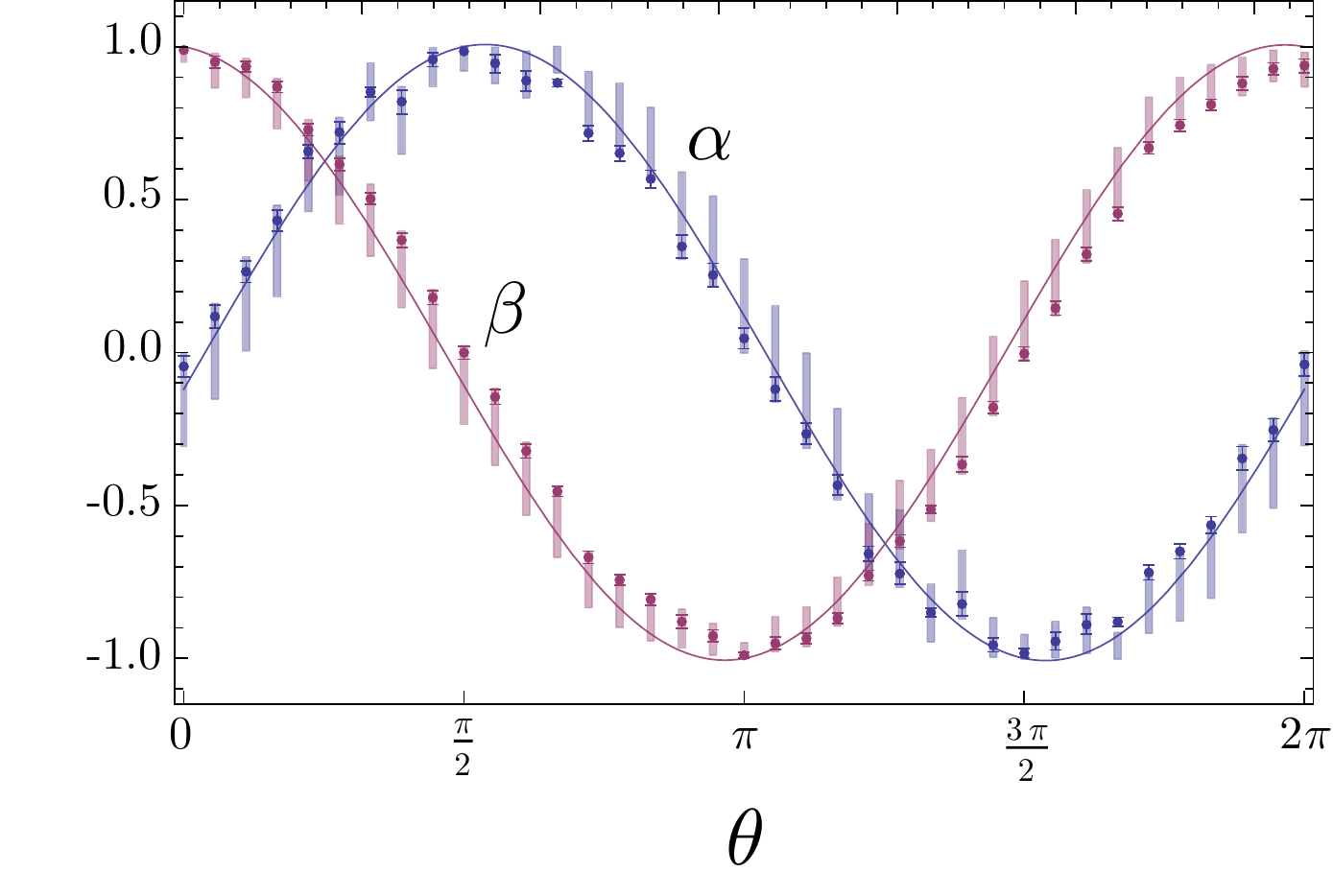}
  \end{center}
\caption{The intervals $[\alpha_{\mathcal M}^{min}, \alpha_{\mathcal M}^{max}]$ and $[\beta_{\mathcal M}^{min}, \beta_{\mathcal M}^{max}]$ obtained using the weak measurement method, for various orientations $\theta$ of the measurement apparatus, are shown as shaded vertical line segments. The smaller intervals of $\alpha_{\mathcal M}$ and $\beta_{\mathcal M}$ obtained when using the combined set $T^{exp}$ are shown as solid line segments. Note, however, that they are dominated by the error bars indicating $1\sigma$ statistical errors and not visible on the scale of this plot. We observe good agreement with the theoretical predictions, shown as solid curves, corresponding to the characterization of the prepared states.}
  \label{fig:Supp_AlphaBeta_Weak}
\end{figure}

While virtually any 2-qubit state would lead to non-trivial bounds in~\eqref{eq:alpha_min_max_T}, the ideal-case analysis suggests that, in order to restrict $\alpha_{\mathcal M}$ to a small interval, the prepared state $\rho_{12}^A$ should be as close as possible to $\rho_{12}^{A,th}$. An analogous analysis applies to $\beta_\mathcal{M}$, using the state $\rho_{12}^B$.
In the experiment we estimated the average values $\moy{{\mathcal M}{\otimes}\sigma_z}_{\rho_{12}^{A}}^{exp}, \moy{{\mathcal M}}_{\rho_{1}^{A}}^{exp}, \moy{{\mathcal M}{\otimes}\sigma_z}_{\rho_{12}^{B}}^{exp}$ and $\moy{{\mathcal M}}_{\rho_{1}^{B}}^{exp}$ for a series of orientations $\theta{\in}[0,2\pi ]$ of the measurement apparatus (Fig.~\ref{fig:Supp_ExpVal_Weak}). The obtained intervals for $\alpha_{\mathcal{M}}$ and $\beta_\mathcal{M}$ are illustrated as line segments in Fig.~\ref{fig:Supp_AlphaBeta_Weak}, including $1\sigma$ statistical errors. 
Importantly, the ranges of possible values of $(\mu, \vec m)$ and therefore of $\alpha_{\mathcal M}$ and $\beta_{\mathcal M}$ can be significantly reduced if we, instead of treating them separately, take into account experimental data from both the weak measurements of $A$ and $B$. We would then estimate $(\mu,\vec m)$ for both $\ve_\mathcal{A}$ and $\ve_\mathcal{B}$ from the set
\begin{align}
T^{exp} &=& \left\{ \! (\mu, \vec m) \left|
\begin{array}{l}
|\mu| + ||\vec m|| \leq 1 \ \text{ and } \\[2mm]
\left\{
\begin{array}{lcl}
\! \mu \, \rho_{12}^{A,1z} + \vec m \! \cdot \! \vec \rho_{12}^{A,\cdot z} &\!=\!& \moy{{\mathcal M} \! \otimes \! \sigma_z}_{\rho_{12}^{A}}^{exp}  \\[1mm]
\! \mu \, \rho_{12}^{B,1z} + \vec m \! \cdot \! \vec \rho_{12}^{B,\cdot z} &\!=\!& \moy{{\mathcal M} \! \otimes \! \sigma_z}_{\rho_{12}^{B}}^{exp}  \\[1mm]
\! \mu + \vec m \cdot \vec \rho_1 &\!=\!& \moy{{\mathcal M}}_{\rho}^{exp}
\end{array}
\!\!\!\! \right.
\end{array}
\right.
\right\} \! .
\label{eq:def_T}
\end{align}
The corresponding results for $\ve_\mathcal{A}$ and $\ve_\mathcal{B}$ are represented by the smaller line segments in Fig.~\ref{fig:Supp_AlphaBeta_Weak}, with error bars representing $1\sigma$ statistical errors. Note that, while one could in principle estimate $(\mu,\vec m)$ from the combination of $T_{\rho_{12}^{A}}^{exp}$ and $T_{\rho_{12}^{B}}^{exp}$, the resulting set of constraints is in many cases too restrictive to be soluble in the presence of experimental imperfections.
Details on the quality of the experimental state preparation and the implementation of the weak measurement can be found in section~\ref{sec:Supp_Exp_Weak} below, where we also present a more detailed error analysis.

\subsection{Estimating $\ve_{\mathcal A}$}
Once the interval $[\alpha_{\mathcal M}^{min}, \alpha_{\mathcal M}^{max}]$ has been determined, with any of the two experimental techniques, one can bound $\ve_{\mathcal A}$ using~\eqref{eq:epsA2}. In the following, we distinguish the case where the same-spectrum assumption is imposed (i.e. $f(m)$ is restricted to take values $\pm 1$), and the case where it is not imposed (i.e. $f(m)$ can take any real value).

In the first case, where $f(m){=}{\pm}1$, Eq.~\eqref{eq:epsA2} implies
\ba
\ve_{\mathcal A}^2 &=& 2 - 2 \sum_m f(m) \operatorname{Re} \Big[ \Tr  \big[{\operatorname M}_m A \rho \big] \Big].
\ea
If $f$ is constant, i.e.\ if $f(+1){=}f(-1){=}\pm 1$, then $\ve_{\mathcal A}^2 = 2 \mp 2 \moy{A}_\rho$, independently of $\mathbb{M}$.
The interesting case is when $f$ is balanced: $f(+1){=}{-}f(-1){=}\tau$, with $\tau{=}{\pm}1$ (chosen to either minimize or maximize $\ve_{\mathcal A}$), resulting in $\ve_{\mathcal A}^2 = 2 - 2 \, \tau \, \alpha_{\mathcal M}$. From the bounds $\alpha_{\mathcal M}^{min} \leq \alpha_{\mathcal M} \leq \alpha_{\mathcal M}^{max}$, we conclude that $\ve_{\mathcal A} \in [\ve_{\mathcal A}^{min}, \ve_{\mathcal A}^{max}]$, with
\ba
(\ve_{\mathcal A}^{min (max)})^2 & \ = \ & \left\{
\begin{array}{ll}
2 - 2 \, \alpha_{\mathcal M}^{max (min)} & \operatorname{ if } \tau = +1 \\[2mm]
2 + 2 \, \alpha_{\mathcal M}^{min (max)} & \operatorname{ if } \tau = -1
\end{array} \right. \! . \quad
\ea

The intervals $[\ve_{\mathcal A}^{min}, \ve_{\mathcal A}^{max}]$ ($[\ve_{\mathcal B}^{min}, \ve_{\mathcal B}^{max}]$) calculated in this way, correspond to the horizontal (vertical) extensions of the green shaded rectangles in Figures~3 and~4 of the main text for the 3-state and the weak measurement method, respectively. In the latter case, the more precise estimates of $(\ve_{\mathcal A}, \ve_{\mathcal B})$ obtained from exploiting data from both weak measurements are shown as darker rectangles in Fig.~4. All intervals include $1\sigma$ statistical errors from a Monte Carlo sampling of the Poissonian photon-counting statistics. Notably, the restricted intervals obtained from Eq.~\ref{eq:def_T} are typically at least one order of magnitude smaller than the statistical errors, leaving room for further improvement.

Without the same-spectrum assumption, instead, for a fixed configuration of the measurement apparatus (i.e.\ a fixed $\mathbb{M}$), we now aim at choosing some output values $f(m)$ that minimize the inaccuracy $\ve_{\mathcal A}$.
It can be shown~\cite{hall2004pih,branciard2013ete} that the optimal value for $f(m)$ is
\ba
f_{opt}(m = \pm 1) &=& \frac{\operatorname{Re}\big[\Tr [{\operatorname M}_\pm A \rho]\big]}{\Tr [{\operatorname M}_\pm \rho]} = \frac{\moy{A}_\rho \pm \alpha_{\mathcal M}}{2\,\Tr [{\operatorname M}_\pm \rho]} . \qquad 
\label{eq:fopt}
\ea 
If, however, $\alpha_{\mathcal M}$ is not known precisely, but only bounded by $\alpha_{\mathcal M}^{min}$ and $\alpha_{\mathcal M}^{max}$, it is not possible to unambiguously define $f(m){=}f_{opt}(m)$.
Instead, we choose to optimize the range of possible values for $\ve_{\mathcal A}$ by either defining $f({+}1){=}\frac{\moy{A}_\rho + \alpha_{\mathcal M}^{min}}{2\,\Tr [{\operatorname M}_+ \rho]}$ or $f({+}1){=}\frac{\moy{A}_\rho + \alpha_{\mathcal M}^{max}}{2\,\Tr [{\operatorname M}_+ \rho]}$, and either $f({-}1){=}\frac{\moy{A}_\rho - \alpha_{\mathcal M}^{min}}{2\,\Tr [{\operatorname M}_- \rho]}$ or $f({-}1){=}\frac{\moy{A}_\rho - \alpha_{\mathcal M}^{max}}{2\,\Tr [{\operatorname M}_- \rho]}$, choosing the combination that gives the smallest range of possible values for $\ve_{\mathcal A}$ from~(\ref{eq:epsAmin_fa}--\ref{eq:epsAmin_fb}) below.

Once $f(m)$ is defined, we use~\eqref{eq:epsA2} and the fact that $\alpha_{\cal M} \in [\alpha_{\cal M}^{min}, \alpha_{\cal M}^{max}]$ to conclude that $\epsilon_{\cal A} \in [\epsilon_{\cal A}^{min}, \epsilon_{\cal A}^{max}]$, with
\ba
(\ve_{\mathcal A}^{min (max )})^2 &=& 1 + \sum_m f(m)^2 \, \Tr  \big[{\operatorname M}_m \rho \big] - \moy{A}_\rho \sum_m f(m) \nonumber \\[-1mm]
&& \quad \ - [f(+1) - f(-1)] \, \alpha_{\mathcal M}^{max (min )} 
\label{eq:epsAmin_fa}
\ea
if $f(+1) - f(-1) \geq 0$, or
\ba
(\ve_{\mathcal A}^{min (max )})^2 &=& 1 + \sum_m f(m)^2 \, \Tr  \big[{\operatorname M}_m \rho \big] - \moy{A}_\rho \sum_m f(m) \nonumber \\[-1mm]
&& \quad \ - [f(+1) - f(-1)] \, \alpha_{\mathcal M}^{min (max )} 
\label{eq:epsAmin_fb}
\ea
if $f(+1) - f(-1) \leq 0$.

The intervals $[\ve_{\mathcal A}^{min}, \ve_{\mathcal A}^{max}]$ ($[\ve_{\mathcal B}^{min}, \ve_{\mathcal B}^{max}]$) calculated in the latter way (without the same-spectrum assumption) correspond to the horizontal (vertical) extensions of the blue shaded rectangles in Figs.~3 and~4 in the main text. Again, for the weak measurement method we can significantly reduce the extension of these regions by refining the bounds on $(\mu, \vec m)$ using the set $T^{exp}$ in Eq.~\ref{eq:def_T}. The corresponding results comprise the darker rectangles in Fig.~(4). All intervals include $1\sigma$ statistical errors. Importantly, these errors again dominate the smaller intervals obtained from $T^{exp}$.

\begin{figure}
  \begin{center}
\includegraphics[width=\columnwidth]{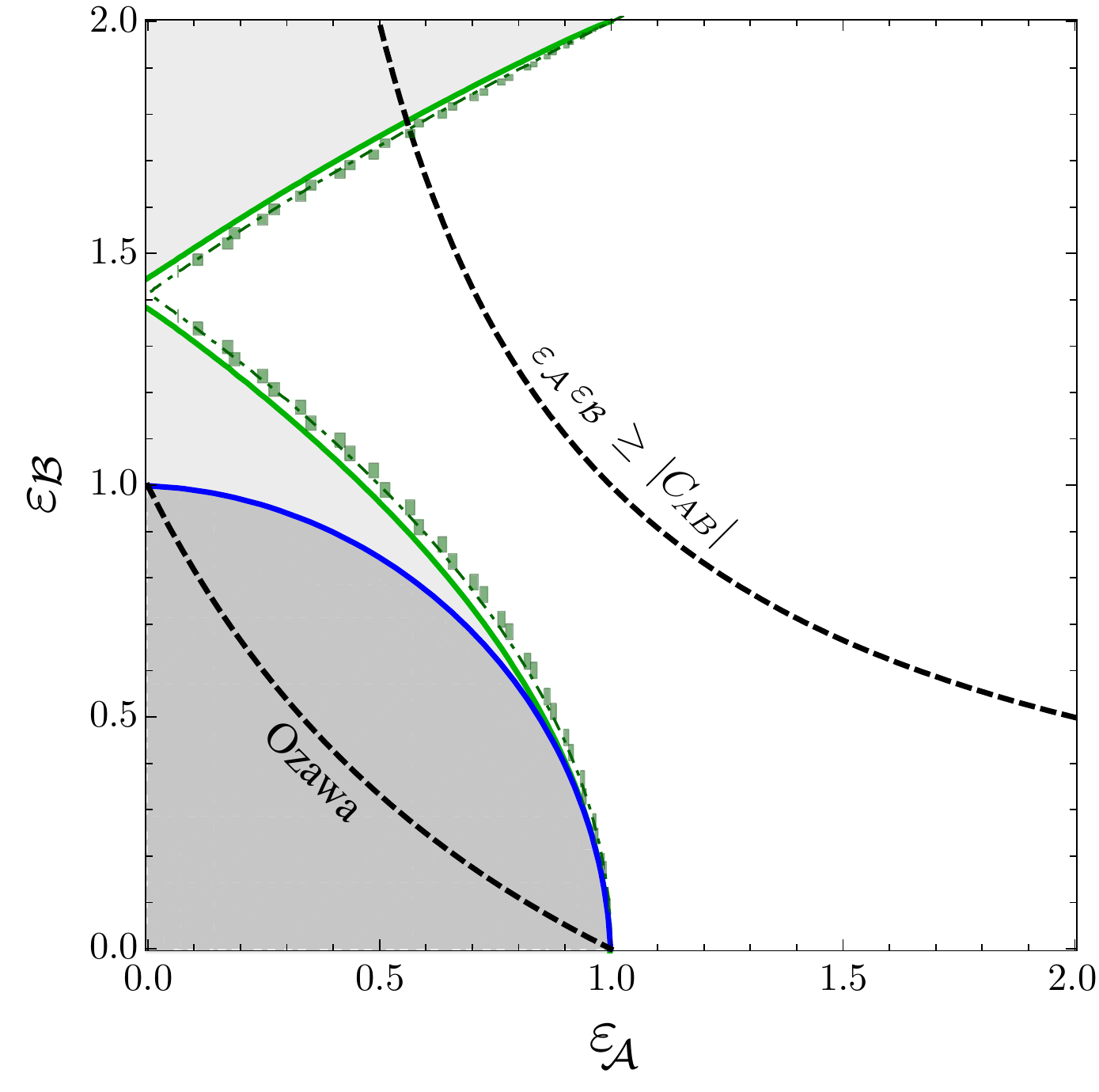}
  \end{center}
\caption{The green rectangles describe the intervals of compatible values for $\ve_\mathcal{A}$ and $\ve_\mathcal{B}$ for the 3-state method, when the same-spectrum assumption is only imposed on ${\mathcal B}$---i.e.\ ${\mathcal B}^2 = B^2 = \one$. The corresponding measurement-disturbance relation is obtained from~(3a) by replacing $\ve_{\mathcal{B}}$ by $\min[1{-}(1{-}\ve_{\mathcal{B}}^2/2)^2, \, \Delta {B}^2]^{1/2}$ and shown as dot-dashed green line for the ideal case. The solid green line represents the bounds imposed by this relation for the experimental values of $\Delta A$,$\Delta B$ and $C_{AB}$. In this case, only $\ve_{\mathcal{B}}$ is upper-bounded. For comparison, the solid blue line shows to the bounds imposed by relation~(3a) as in Fig~(3). All intervals include $1\sigma$ statistical errors.}
  \label{fig:Tomo_epsilons_SameSpectB}
\end{figure}

\begin{figure}
  \begin{center}
\includegraphics[width=\columnwidth]{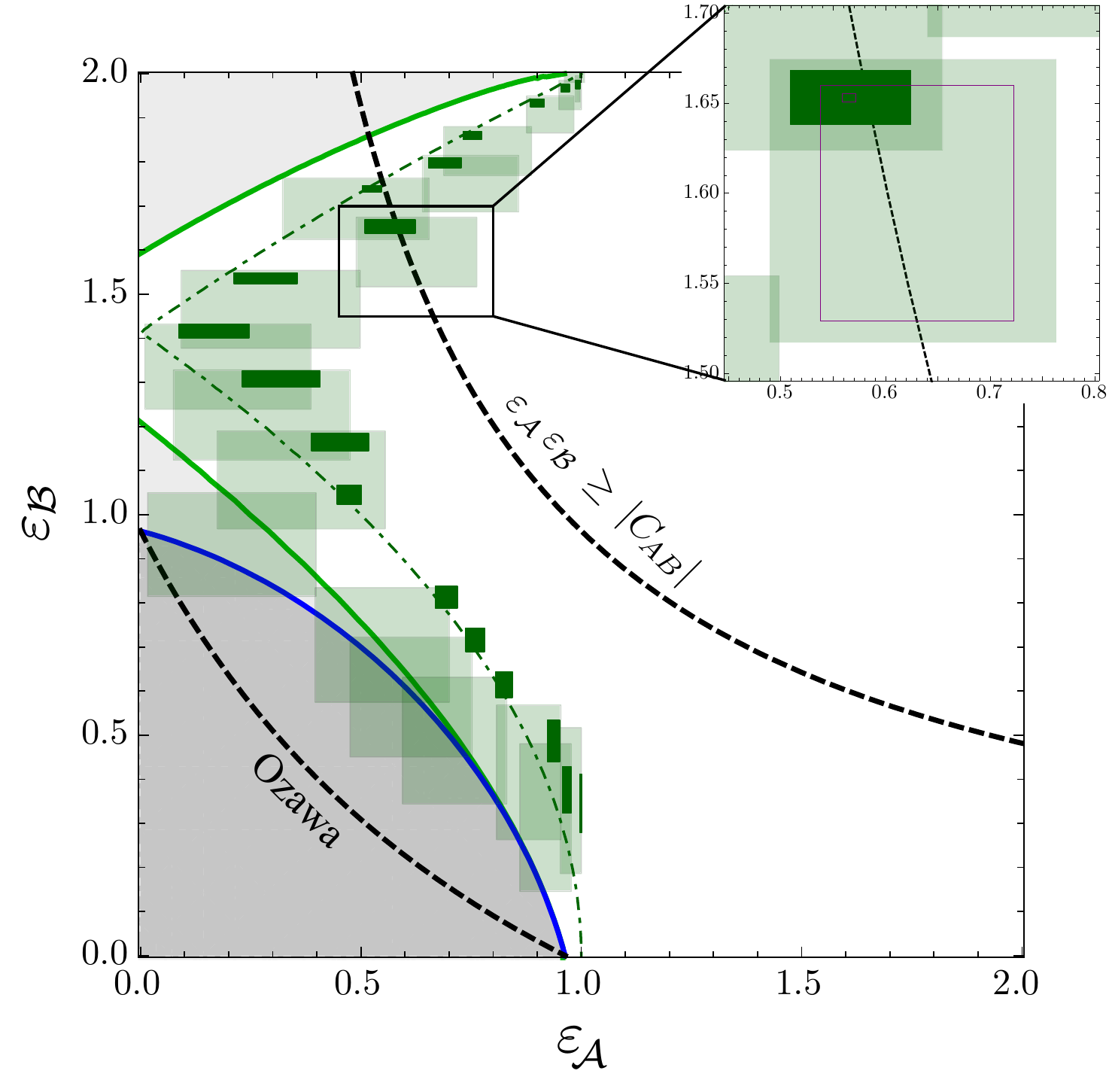}
  \end{center}
\caption{As in Figure~\ref{fig:Tomo_epsilons_SameSpectB}, for the weak measurement method. The darker rectangles correspond to the more precise estimates of $\ve_\mathcal{A}$ and $\ve_\mathcal{B}$ obtained from Eq.~\ref{eq:def_T}. The solid blue line shows relation~(3a) for comparison. All intervals include $1\sigma$ statistical errors. The inset illustrates the size of the obtained intervals (outlined purple rectangles) with respect to the $1\sigma$-range of statistical errors associated with the respective intervals (shaded regions outside the outlined rectangles).}
  \label{fig:Weak_epsilons_SameSpectB}
\end{figure}

While for simplicity and clarity in the plots of the main text the same-spectrum assumption as been imposed on both observables $\mathcal{A}$ and $\mathcal{B}$, one can also consider a situation where the same-spectrum assumption is imposed on only one of them (e.g. on $\mathcal{B}$ only). As discussed in the main text, this is a natural assumption when one interprets our experiment as implementing a measurement-disturbance scenario. The corresponding results are shown in Figs.~\ref{fig:Tomo_epsilons_SameSpectB} and~\ref{fig:Weak_epsilons_SameSpectB} for the 3-state and weak measurement method, respectively.

\section{Experimental details}
\label{Sec:Supp2}
Here we describe in details the characterization of the experimental setup, including the quality of the state preparation and of the implementation of various crucial elements for both experimental approaches.

For both the 3-state and the weak measurement methods, the state preparation was performed by a Glan-Taylor calcite polarizer with a nominal extinction ratio of $100{,}000{:}1$, followed by a quarter-wave plate (QWP) and a half-wave plate (HWP), which were carefully characterized. The same setup in reverse order was used to perform arbitrary projective measurements, required for quantum state tomography and to implement the measurement $\mathcal{M}=\mu\one+\vec m \cdot \vec \sigma$. While this design was specifically chosen to prevent any systematic bias $\mu$, such that $\mathcal{M}$ is as close to a projective measurement as possible, it only performs a one-outcome measurement. The binary observable $\mathcal{M}$ is therefore realized by measuring both projectors individually and combining the results appropriately. We used high precision motorized wave plates to ensure the accurate and repeatable implementation of these measurements.

Due to the linear design of the experiment we were able to calibrate the employed wave plates \emph{in-situ} using single photons, avoiding any additional systematic errors arising from external calibration. We find the optical axis of each wave plate with a typical precision of $10^{-2}$, limited only by manufacturing imperfections and, to a small part, by the Poissonian statistics of the single-photon source. We also characterize the actual retardance $r$ of each wave plate through its relation to the visibility of the measured fringes:  $V{=}\frac{2r}{\lambda}$. Using very high quality optical elements, we still find relative deviations from the nominal retardance on the order of $10^{-3}$ or up to $\lambda/200$; see Tables~\ref{tab:tomo_WPdata} and~\ref{tab:weak_WPdata}.

\begin{table}[h!]
\begin{tabular}{c@{\hskip 2em} c@{\hskip 2em} c}
\hline\hline
Element & $\delta\phi$ & $V$ \\
\hline
QWP1 & $0.0566$ & $0.4986(12)$ \\
HWP1 & $0.0203$ & $0.9939(9)$ \\
HWP2 & $0.0210$ & $0.9943(9)$ \\
QWP2 & $0.0579$ & $0.4859(12)$ \\
\hline\hline
\end{tabular}
\caption{Shown are the standard deviation $\delta\phi$ from the fit of the optical axis, as well as the visibility of the observed fringes in the measured coincidence counts for all wave plates used in the 3-state method.}
\label{tab:tomo_WPdata}
\end{table}

\begin{table}[h!]
\begin{tabular}{c@{\hskip 2em} c@{\hskip 2em} c}
\hline\hline
Element & $\delta\phi$ & $V$ \\
\hline
QWP1 & $0.0330$ & $0.5017(7)$ \\
HWP1 & $0.0096$ & $0.9961(4)$ \\
QWP2 & $0.0412$ & $0.5026(8)$ \\
HWP2 & $0.0263$ & $0.9959(12)$ \\
\rule{0pt}{1.5em}HWP3 & $0.0154$ & $0.9958(6)$ \\
QWP3 & $0.0253$ & $0.5028(12)$ \\
HWP4 & $0.0116$ & $0.9957(5)$ \\
QWP4 & $0.0114$ & $0.4981(20)$ \\
\hline\hline
\end{tabular}
\caption{Shown are the standard deviation $\delta\phi$ from the fit of the optical axis, as well as the visibility $V$ of the observed fringes in the measured coincidence counts for the preparation (1,2) and measurement (3,4) wave plates used in the weak measurement method.}
\label{tab:weak_WPdata}
\end{table}

From a theoretical model only minor reductions in the fidelity of the implemented projective measurements and single-qubit unitaries are expected due to these imperfections. We note, however, that they still lead to observable deviations in the experimental data, showing the high degree of sensitivity of this experiment. While the uncertainty in the optical axis of the wave plates is manifest as a tilt of the plane of measurement in the Bloch sphere, the imperfect retardance causes more complicated behavior with no obvious interpretation on the Bloch sphere.

As an example to illustrate the magnitude of these errors, we analyze the implementation of the projective measurement $\mathcal{M}$, finding a tilt of the measurement plane of $0 \pm 0.0697(6)^\circ$ and $0\pm 0.0423(4)^\circ$ for the 3-state and weak measurement methods, respectively.
Taking both errors into account we still expect a process fidelity of $\mathcal{F}_p=0.99996(1)$ for the 3-state and $\mathcal{F}_p=0.999998(3)$ for the weak measurement method. Nevertheless, the corresponding deviation from the $xz$-plane covers a significant range of $[0^\circ,1.03^\circ]$ and $[0^\circ,0.247^\circ]$ averaging at $0.66(11)^\circ$ and $0.158(85)^\circ$ for the two techniques. Our treatment of the measurement apparatus as a black box, however, makes the data analysis insensitive to the above errors in the actual implementation of $\mathcal{M}$.

To avoid any systematic errors in the state preparation we perform a fine-grained search over the respective wave plate angles until optimal state parameters are obtained. As figures of merit in this optimization for the 3-state method we considered the fidelity of the state $\rho$ with $\ketbra{{+}y}$ as well as, in particular, the values of $\moy{A}_{\rho}$ and $\moy{B}_{\rho}$.
For the weak measurement method we took into account in particular the components of the Bloch vectors $\vec \rho_{12}^{A/B,\cdot z}$ and $\vec \rho_{1}^{A/B}$, since they encompass most of the quality factors of the prepared state.

Using this calibrated setup we carefully characterized the experimental state preparation for both methods. All the errors in the following are based on a Monte-Carlo simulation of the Poissonian counting statistics of the single-photon source.

\subsection{The 3-state method}
\label{sec:Supp_Exp_Tomo}
The reconstructed density matrix of the experimental state $\rho\simeq \ketbra{{+}y}$ is
\begin{equation}
\begin{pmatrix} 0.5109(1) & -0.0172(1) - 0.499157(7) i \\  -0.0172(1) + 0.499157(7) i & 0.4891(1) \end{pmatrix} \! .
\end{equation}

Further characteristics of this state are summarized in Table~\ref{tab:tomo_rhodata}.
\begin{table}[h!]
\begin{tabular*}{\columnwidth}{@{\extracolsep{\stretch{1}}}*{5}{c}}
\hline\hline
$\mathcal{F}$ & $\mathcal{P}$ & $\moy{A}_\rho$ & $\moy{B}_\rho$ & $C_{\!AB}^2$ \\
\hline
$0.999172(7)$ & $0.99917(2)$ & $-0.0344(2)$ & $0.0216(2)$ & $0.99669(3)$\\
\hline\hline
\end{tabular*}
\caption{Characteristic data for the state $\rho$ on which the joint measurement is to be approximated.}
\label{tab:tomo_rhodata}
\end{table}

In addition to the state $\rho$, the 3-state method requires the preparation of 2 additional states $\rho_1 \simeq A\rho A$ and $\rho_2 \simeq (\one+A)\rho(\one+A)$ for the estimation of $\ve_\mathcal{A}$ and similarly $\rho_1'$ and $\rho_2'$ for the estimation of $\ve_\mathcal{B}$. Note, that $\rho_1=\rho_1'$ experimentally, but due to imperfections in $\rho$, they correspond to slightly different theoretical states $A\rho A$ and $B\rho B$, respectively. Some relevant characteristic data for these states can be found in Table~\ref{tab:tomo_statedata}.

\begin{table}[h!]
\begin{tabular*}{\columnwidth}{@{\extracolsep{\stretch{1}}}*{5}{c}}
\hline\hline
 & $\rho_1$ & $\rho_2$ & $\rho_1'$ & $\rho_2'$ \\
\hline
$\mathcal{F}$ & $0.999315(6)$ & $0.999533(5)$ & $0.999293(6)$ & $0.999579(5)$ \\
$\mathcal{P}$ & $0.99761(2)$ & $0.99916(1)$ & $0.99761(2)$ & $0.99940(1)$ \\
$\Delta\varphi [^\circ]$ & $5.24(1)$ & $2.48(1)$ & $5.27(1)$ & $2.35(1)$\\
\hline\hline
\end{tabular*}
\caption{Characteristic data for the additional states $\rho_1^{(\prime)}$ and $\rho_2^{(\prime)}$ used in the 3-state method. Here $\Delta\varphi$ refers to the angular deviation of the state vector on the Bloch sphere when compared to the ideal state. The fidelities and angular deviations are calculated with respect to the states $A\rho A$, $(\one+A)\rho(\one+A)$ etc., for the carefully characterized experimental state $\rho$.}
\label{tab:tomo_statedata}
\end{table}

For the projective measurements performed for quantum state tomography we obtain typical process fidelities of $\mathcal{F}_p{=}0.99978(4)$ and an average angle deviation of $1.3(1)^\circ$. We note that the errors are dominated by the imperfect retardance of the QWP2 in Table~\ref{tab:tomo_WPdata}, which causes an angle deviation of as much as $2.7(1)^\circ$ in the $\sigma_y$ measurement, while not affecting the $\sigma_x$ and $\sigma_z$ measurements. This translates into an average error in $\moy{\sigma_y}$ of $0.0137(2)$ and in a relatively large rotation of the above Bloch vectors from their ideal directions (see Table~\ref{tab:tomo_statedata}). Recalling that the Bloch vectors characterize the set $S_{\rho,\rho_1,\rho_2}^{exp}$ in~\eqref{eq:def_Srhos}, it becomes obvious that any errors in the state tomography directly affect the size of this set and thus the ranges of compatible values for $\alpha_{\mathcal{M}}$ and $\beta_{\mathcal{M}}$.

Importantly, systematic errors in the state tomography and the resulting potential mismatch between the experimental state (on which $\moy{\mathcal M}_{\rho}$ is measured) and the reconstructed state (from which the Bloch vectors are calculated) can invalidate the equality in the defining equations in~\eqref{eq:def_Srhos}. This in turn might result in the set $S_{\rho,\rho_1,\rho_2}^{exp}$ containing no physical solutions (i.e.\ such that $|\mu|+\|\vec m\|\leq 1$ is satisfied) for some sets of experimental data. This problem can be overcome by performing a Monte-Carlo routine based on the errors obtained from the characterization of the setup.

\subsection{The weak measurement method}
\label{sec:Supp_Exp_Weak}
The crucial element of the optical setup for this method is the partially polarizing beam splitter (PPBS) implementing a controlled-phase gate~\cite{Langford2005}. A nominal reflection (transmission) of $R_V=2/3$ ($T_V=1/3$) for vertically polarized light ($\ket{V}$) and $R_H=0$ ($T_H=1$) for horizontally polarized light ($\ket{H}$) ensures optimal gate operation with a success probability of $1/9$. Whilst taking great care in the initial alignment of this PPBS with respect to the initial polarization reference set by the 4 Glan-Taylor calcite polarizers, the polarization extinction ratio is still reduced to about $5{,}600{:}1$ ($8{,}600{:}1$) in the signal (meter) arm due to a slight polarization rotation in the material. We further measured a splitting ratio of $\frac{R_V}{T_V} = 2.02(3)$ ($1.97(3)$) with a residual reflection of horizontally polarized light of $0.00459(6)$ ($0.00333(5)$) in the signal (meter) arm. From a theoretical model based on these parameters we still expect a process fidelity of $\mathcal{F}_p{=}0.993(3)$ for the implementation of the \textsc{cnot} operation. Importantly, we were able to reduce the number of optical elements and thus potential sources of error by replacing the additional PPBS, usually required for amplitude compensation by a ``pre-biased'' input---i.e.\ instead of reducing the amplitude of $\ket{H}$ after the gate, we increase the amplitude of $\ket{V}$ before the gate accordingly.

Comparing the above estimated process fidelity to the characterization of the gate via process tomography, summarized in Table~\ref{tab:weak_processtomo}, suggests that the manufacturing deviations of the main PPBS are not the limiting factor in the quality of our implementation. We believe that the imperfections are to a large part caused by polarization-dependent loss in the PPBS, as well as mode mismatch and residual frequency-correlations of the single photons. This is supported by a more detailed analysis, showing decreased relative (with respect to the ideal case for this PPBS) interference visibility of typically $\overline{V_r}{=}0.958(9)$ for input states and projective measurements different from $\ketbra{VV}$, for which the gate is typically optimized (achieving $V_r{=}1.00(1)$ in our case). This is again attributed to polarization dependent loss in the PPBS, but also to steering of the two beams, caused by the various wave plates, reducing the spatial overlap.

\begin{table}[h!]
\begin{tabular*}{\columnwidth}{@{\extracolsep{\stretch{1}}}*{5}{c}}
\hline\hline
$\mathcal{F}_p$ & $\mathcal{P}_p$ & $\overline{\mathcal{F}}$ & $\overline{\mathcal{P}}$ & $V_r$ \\
\hline
$0.964(1)$ & $0.931(3)$ & $0.98(1)$ & $0.97(2)$ & $1.00(1)$\\
\hline\hline
\end{tabular*}
\caption{Characteristic data of the \textsc{cnot} gate obtained via process tomography. Here $\overline{\mathcal{F}}$ and $\overline{\mathcal{P}}$ denote the average fidelity and purity of the tomographic set of states, respectively. The errors on $\mathcal{F}_p$ and $\mathcal{P}_p$ result from a Monte Carlo simulation of the Poissonian counting statistics, while the errors on the other parameters represent the standard deviations of the corresponding sets of measurements.
}
\label{tab:weak_processtomo}
\end{table}

We also find that the PPBS imparts a birefringent phase between the $\ket{H}$ and $\ket{V}$ components of the input state of $0.376(9)$ and $0.384(17)$ in the signal and meter arms, respectively. This has been taken into account and compensated by adapting the state preparation appropriately. While insufficient compensation of this effect might reduce the fidelity of the operation, it does not affect quality of the 2-photon interaction and thus cannot decrease the gate purity. We find, however, a rather weak dependence on this phase compared to other sources of error.

From the measured wave plate errors we further estimate a process fidelity of $\mathcal{F}_p{=}0.999977(9)$ for the implementation of the final Hadamard gate on the meter qubit, corresponding to a $1\sigma$ angle-deviation in the rotation axis of $0.017^\circ$. Note, however, that in order to reduce the number of optical elements and thus potential sources of error, the Hadamard transformation was incorporated in the final projective $\sigma_z$ measurement. For this joint operation we achieve a process fidelity of $\mathcal{F}_p{=}0.999988(5)$. While, judging from these values we would expect a very low error rate, careful analysis reveals a systematic deviation of the direction of the projective measurement on the Bloch Sphere of as much as $0.39(9)^\circ$.
Similarly, the Hadamard gate in the meter arm before the \textsc{cnot} (expected process fidelity of $\mathcal{F}_p{=}0.99996(3)$ and angle-deviation $0.026^\circ$) has been included in the state preparation in an effort to reduce the number of optical elements.

\begin{figure}
  \begin{center}
\includegraphics[width=\columnwidth]{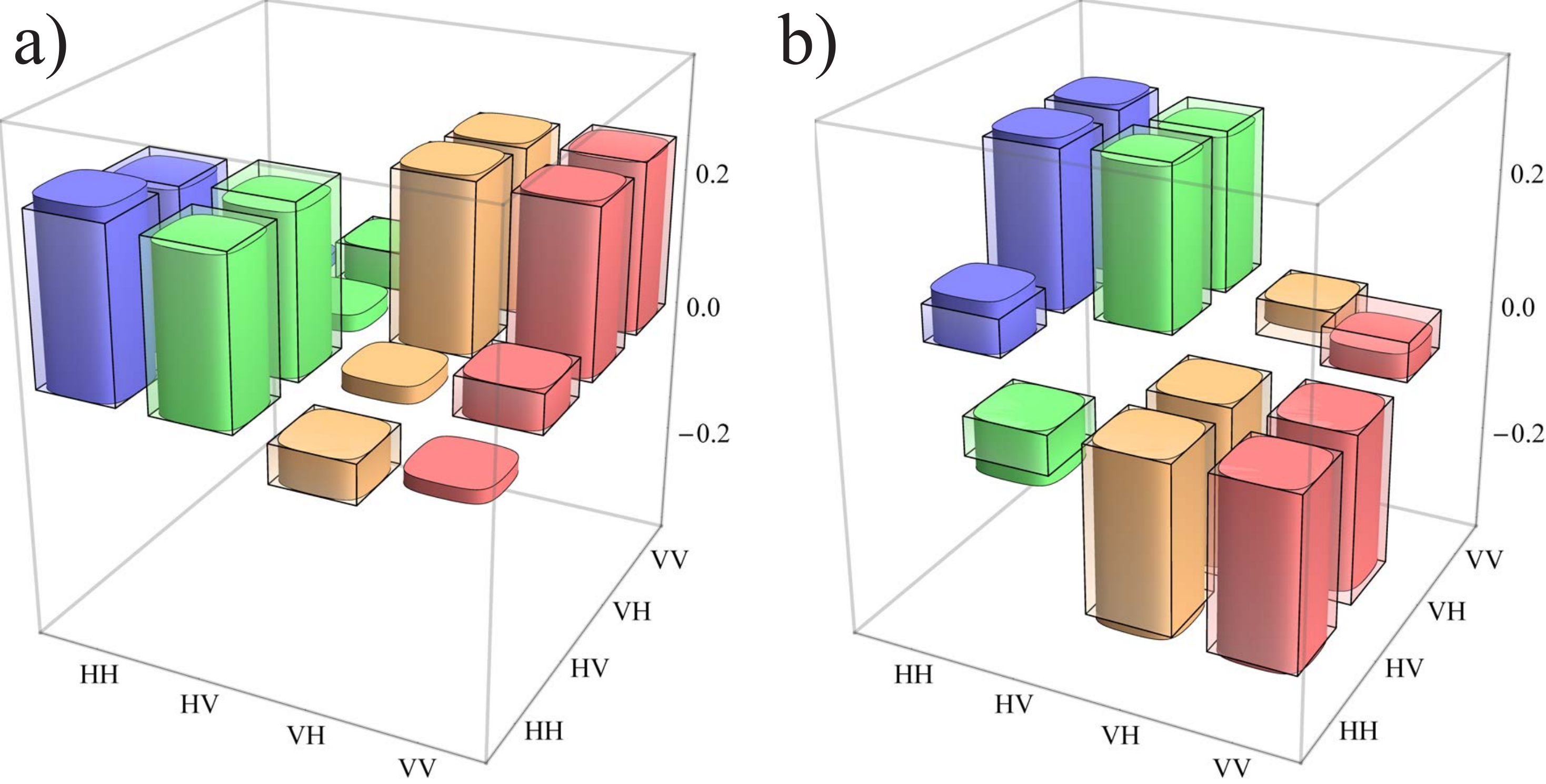}
  \end{center}
\caption{Shown are the a) real and b) imaginary part of the reconstructed density matrix for the state $\rho_{12}^A$. The wire-frame represents the ideal density matrix given by Eq.~\eqref{eq:rho12Ath} for the corresponding value of $\kappa=-0.226(4)$.}
  \label{fig:DMrho12A}
\end{figure}

\begin{figure}
  \begin{center}
\includegraphics[width=\columnwidth]{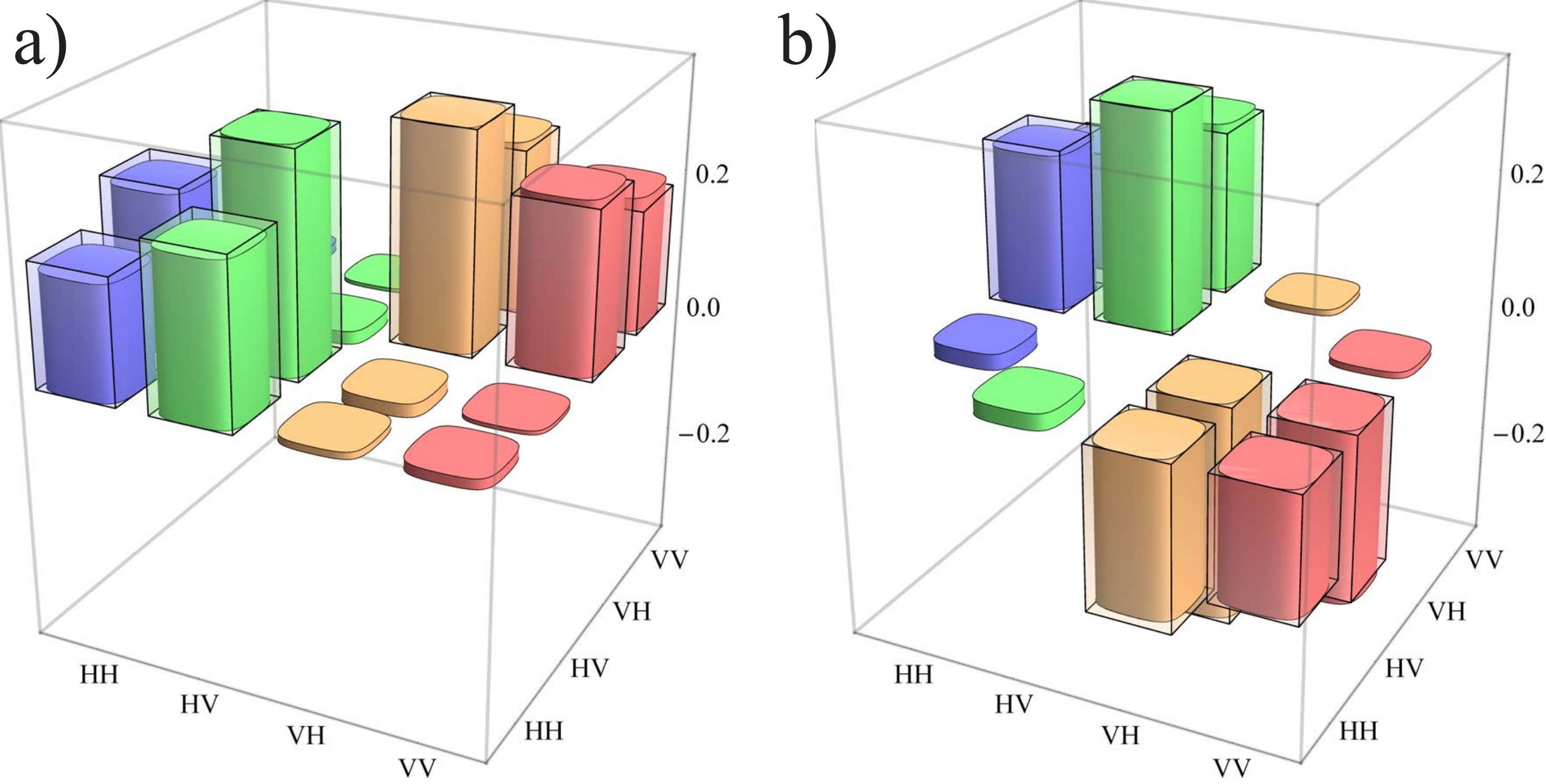}
  \end{center}
\caption{Shown are the a) real and b) imaginary part of the reconstructed density matrix for the state $\rho_{12}^B$. The wire-frame represents the ideal density matrix given by Eq.~\eqref{eq:rho12Bth} for the corresponding value of $\kappa=-0.280(2)$.}
  \label{fig:DMrho12B}
\end{figure}

As discussed above, we performed an optimization of the wave plate angles to improve the quality of the initial state preparation. Using this technique, we are able to achieve very high quality in the state preparation of both the meter qubit and the state $\rho$ onto which the joint measurement of $A$ and $B$ is approximated. Typical fidelities for the preparation of the meter state are $\mathcal{F}{=}0.99976(8)$ with a purity of typically $\mathcal{P}{=}0.9997(2)$. The reconstructed density matrix of the experimental state $\rho = \frac{1}{2} (\rho_1^A + \rho_1^B)$ of the first qubit after the 2-qubit interactions is
\begin{equation*}
\begin{pmatrix} 0.493(2) & 0.001(2) - 0.4804(6) i \\ 0.001(2) + 0.4804(6) i & 0.507(2) \end{pmatrix} ,
\end{equation*}
with more details on the characteristics summarized in Table~\ref{tab:weak_rhodata}. The corresponding 2-qubit density matrices $\rho_{12}^A$ and $\rho_{12}^B$ after the interaction allowing the semi-weak measurements of $A$ and $B$, respectively, are shown in figures~\ref{fig:DMrho12A} and~\ref{fig:DMrho12B}. These figures also include the ideal states $\rho_{12}^{A,th}$ and $\rho_{12}^{B,th}$ of Eqs.~\eqref{eq:rho12Ath} and~\eqref{eq:rho12Bth} for the respective values of $\kappa=-0.226(4)$ and $\kappa=-0.280(2)$.

\begin{table}[h!]
\begin{tabular*}{\columnwidth}{@{\extracolsep{\stretch{1}}}*{5}{c}}
\hline\hline
$\mathcal{F}$ & $\mathcal{P}$ & $\moy{A}_\rho$ & $\moy{B}_\rho$ & $C_{\!AB}^2$ \\
\hline
$0.99998(5)$ & $0.964(1)$ & $0.001(4)$ & $-0.007(4)$ & $0.928(2)$\\
\hline\hline
\end{tabular*}
\caption{Characteristic data for the state $\rho$ on which the joint measurement is to be approximated in the weak measurement method.}
\label{tab:weak_rhodata}
\end{table}

Analyzing the quality of the projective measurements used for state tomography we find an average process fidelity of $\mathcal{F}_p{=}0.99998(1)$ and $\mathcal{F}_p{=}0.99988(2)$ for the meter and signal arm, respectively. The corresponding angular deviation of the projective measurements on the Bloch sphere averages to $0.4(1)^\circ$ and $0.8(1)^\circ$, respectively. While in the meter arm this systematic deviation is similar for all states in the tomographic set, the error in the signal arm again seems to be dominated by the measurement quarter-wave plate, resulting in a deviation of $2.1(2)^\circ$ for the $\sigma_y$. In an effort to quantify the effect to the above errors, we investigate the overlap between the states $\rho_1^A$ and $\rho_1^B$, measured after different 2-qubit interactions, finding a fidelity of $\mathcal{F}{=}0.9992(1)$.

As discussed in section~\ref{Sec:Supp_Alpha_Weak} the parameters describing the set $T_{\rho_{12}^{A}}^{exp}$ in~\eqref{eq:def_TrhoA} are very sensitive to experimental imperfections. In addition, the contrast in the expectation value $\moy{{\mathcal M} \otimes \sigma_z}_{\rho_{12}^{A,th}}$ as well as in the components of the Bloch vector $\vec \rho_{12}^{A,\cdot z}$ decreases proportionally to the measurement strength $|\kappa|$. Furthermore, the mismatch between the experimental state and the reconstructed state, as discussed above, is even bigger in the case of a two-qubit reconstruction. These imperfections directly affect the defining constraints of the set $T_{\rho_{12}^{A}}^{exp}$ and thus also the size of the ranges of compatible values for $\alpha_{\mathcal{A}}$, especially for low measurement strengths. This might also cause, in particular, the more restrictive set $T^{exp}$ obtained when combining the data from both the weak measurements of $A$ and $B$, to be unphysical. We find, however, that with our experimental setup $T^{exp}$ remains non-empty for all tested cases and can therefore be used to obtain more precise estimates of $\ve_\mathcal{A}$ and $\ve_\mathcal{B}$. Also for this case we performed a Monte-Carlo routine to estimate the errors caused by the Poissonian counting statistics.


\begin{thebibliography}{28}
\bibitem[{NoteS2()}]{NoteS2}
\bibinfo{note}{for more confidence, or for generalizing the technique to
  larger dimensions, note that it can also be cast as a semi-definite
  programming problem, which can be solved efficiently~\cite{Vandenberghe:1996fk}.}
\end{thebibliography}
\end{document}